\documentclass[a4paper,fleqn,usenatbib]{mnras}

\usepackage[T1]{fontenc}
\usepackage{aecompl}

\usepackage{graphicx}	
\usepackage{amsmath}	
\usepackage{amssymb}	

\usepackage{xspace}
\usepackage{siunitx}
\usepackage{tikz}
\usepackage{mhchem}

\usepackage{multirow}

\newcommand{\snia}{SN~Ia}

\newcommand{\snias}{SNe~Ia}
\newcommand{\sniasx}{SNe~Iax}
\newcommand{\stella}{\textsc{Stella}\xspace}
\newcommand{\artis}{\textsc{Artis}\xspace}
\newcommand{\sedona}{\textsc{Sedona}\xspace}
\newcommand{\nuc}[2]{\ensuremath{\mathrm{^{#1}#2}}}

\graphicspath{{./}}

\title[Early SNe Ia light curves]{Early light curves for Type Ia supernova explosion models}

\author[Noebauer~et~al.]{U.~M.~Noebauer,$^1$\thanks{unoebauer@mpa-garching.mpg.de}
M.~Kromer,$^{2,3}$
S.~Taubenberger,$^{1,4}$
P.~Baklanov,$^{5,6,7}$
S.~Blinnikov,$^{5,8,9}$
\newauthor
E.~Sorokina,$^{5,8,10}$
and W.~Hillebrandt$^{1}$\\
$^1$Max-Planck-Institut f\"ur Astrophysik, Karl-Schwarzschild-Str.~1, D-85741 Garching, Germany\\
$^2$Zentrum f\"ur Astronomie der Universit\"at Heidelberg, Institut f\"ur Theoretische Astrophysik, Philosophenweg 12, D-69120 Heidelberg, Germany\\
$^3$Heidelberger Institut f\"ur Theoretische Studien, Schloss-Wolfsbrunnenweg 35, D-69118 Heidelberg, Germany\\
$^4$European Southern Observatory, Karl-Schwarzschild-Str.~2, D-85748 Garching, Germany\\
$^5$Institute for Theoretical and Experimental Physics (ITEP), 117218 Moscow, Russia\\
$^6$Novosibirsk State University (NSU), Novosibirsk 630090, Russia\\
$^7$National Research Nuclear University (MEPhI), Moscow 115409, Russia\\
$^8$Kavli Institute for the Physics and Mathematics of the Universe (WPI), The University of Tokyo, Kashiwa, Chiba 277-8583, Japan\\
$^9$All-Russia Research Institute of Automatics (VNIIA), 127005 Moscow, Russia\\
$^{10}$Sternberg Astronomical Insitute, M.V.Lomonosov Moscow State University,
119234 Moscow, Russia
}

\date{Accepted XXX. Received YYY; in original form ZZZ}

\pubyear{2017}

\begin{document}
\label{firstpage}
\pagerange{\pageref{firstpage}--\pageref{lastpage}}
\maketitle

\begin{abstract}
  Upcoming high-cadence transient survey programmes will produce a wealth of
  observational data for Type Ia supernovae. These data sets will contain
  numerous events detected very early in their evolution, shortly after
  explosion. Here, we present synthetic light curves, calculated with the
  radiation hydrodynamical approach \textsc{Stella} for a number of different
  explosion models, specifically focusing on these first few days after
  explosion. We show that overall the early light curve evolution is similar
  for most of the investigated models.  Characteristic imprints are induced by
  radioactive material located close to the surface. However, these are very
  similar to the signatures expected from ejecta--CSM or ejecta--companion
  interaction. Apart from the pure deflagration explosion models, none of our
  synthetic light curves exhibit the commonly assumed power-law rise. We
  demonstrate that this can lead to substantial errors in the determination of
  the time of explosion. In summary, we illustrate with our calculations that
  even with very early data an identification of specific explosion scenarios
  is challenging, if only photometric observations are available. 
\end{abstract}




\section{Introduction}
\label{sec:intro}

Type Ia supernovae (SNe~Ia) are believed to originate from thermonuclear
explosions in carbon--oxygen (CO) white dwarfs \citep[WDs;][]{Hoyle1960}.
However, details of the progenitor stars and explosion mechanisms are still
uncertain and various models are discussed. One uncertainty concerns the mass
of the exploding object, in particular whether the explosion occurs in a WD
close to the Chandrasekhar mass limit or well below.  Furthermore, the details
of the thermonuclear burning, i.e.\ whether the flame proceeds as a
deflagration, a detonation or a mixture of both, are still unresolved
\citep{Hillebrandt2000}. It is also still heavily debated whether the WD
accretes material from a companion star or if the supernova occurs in the
merger with a second WD \citep[e.g.][]{Ruiter2009}.  Finally, more exotic
models, such as head-on collisions \citep[e.g.][]{Rosswog2009} or the
core-degenerate mechanism \citep{Soker2014}, are discussed as well. For a
comprehensive overview on the open questions, see the recent reviews by
\citet{Hillebrandt2013} or \citet{Maoz2014}.

Robotic transient searches, such as the Panoramic Survey Telescope And Rapid
Response System (Pan-STARRS) or the Palomar Transient Factory (PTF), conducted
during the last decade have presented the scientific community with a wealth of
supernova observations.  Since the cadence of these surveys continuously
shortened, supernovae were caught at ever earlier times, shortly after
explosion \citep[e.g.\ SN~2011fe,][]{Nugent2011}. This situation will again
dramatically improve with upgrades of current surveys, for example, of the
All-sky Automated Survey for SuperNovae (ASAS-SN), and with future transient
searching campaigns, such as the Zwicky Transient Facility or the Large
Synoptic Survey Telescope, which will drastically increase the number of
\snias{} detected in their early evolutionary phases. 

Entirely new avenues to study supernova physics and learn more about the
progenitor system, the circumstellar environment in the immediate vicinity of
the explosion site and even the actual explosion mechanism are opened by such
data sets. These prospects have sparked increased interest in examining the
expected very early observational signatures from a theoretical perspective
\citep[e.g.][]{Dessart2014b}. In particular, the signatures from cooling of
shock heated ejecta \citep{Piro2010, Rabinak2012} or the characteristic
imprints in the early light curve induced by the interaction of ejecta with a
companion \citep{Kasen2010} or with circumstellar material \citep[CSM,
e.g.][]{Piro2016} have been studied.  Likewise, the distribution of radioactive
material in the outer ejecta regions is expected to leave traces in the early
observables as well \citep[e.g.][]{Diehl2014, Piro2014, Piro2016}.

Some of the diagnostic possibilities that very early observations offer have
already been showcased by a number of individual, nearby objects. For example,
thanks to the availability of observations shortly after explosion,
\citet{Nugent2011} and \citet{Bloom2012} were able to constrain the progenitor
radius of SN~2011fe to unprecedented precision. Similar studies have been
attempted for other \snias{}, among them SN~2013dy \citep{Zheng2013},  SN~2014J
\citep{Zheng2014, Goobar2015}, ASASSN-14lp \citep{Shappee2016} and SN~2015F
\citep{Im2015}, for which very early and densely sampled observational data
have been collected. Although no companion interaction has been detected in
past surveys \citep{Hayden2010a, Bianco2011}, the observation of such
signatures has recently been claimed for a few individual objects, in
particular for the normal SN~2012cg \citep{Marion2016} and the sub-luminous
iPTF14atg \citep{Cao2015}.  These observations are typically interpreted as
evidence for the single degenerate scenario for \snias{} (but see e.g.\
\citealt{Livio2003}, \citealt{Raskin2013}, \citealt{Liu2016},
\citealt{Kromer2016} and \citealt{Shappee2016a}). 

With this work, we aim at exploring the very early observables for a set of
detailed explosion models of \snias{}. We calculate synthetic light curves in
various photometric bands for these models with the well-established radiation
hydrodynamics method \stella \citep{Blinnikov1993, Blinnikov1998,
Blinnikov2000} for the first ten days after explosion. We study the overall
shape and the rise behaviour of these early light curves and search for
characteristic signatures that may help to robustly discriminate different
explosion mechanisms based on early photometric data. Our work is organized as
follows. We briefly review the key aspects of the numerical approach, we rely
on in Section~\ref{sec:method} and present the various explosion models we
investigate in Section~\ref{sec:models}. The early light curves calculated with
\stella for these models are presented in Section~\ref{sec:results} and
discussed in detail in Section~\ref{sec:discussion}. We conclude with a summary
and an outlook about future work in Section~\ref{sec:summary}.

\section{Method}
\label{sec:method}

During the early phases of supernova evolution, the ejecta are still very dense
and thus optically thick. Radiation has to diffuse from nickel-rich zones where
it is generated to the very outer ejecta regions which are transparent. The
high optical thickness of the ejecta during the early phases argues against the
use of Monte Carlo-based radiative transfer approaches such as \artis
\citep{Kromer2009} or \sedona \citep{Kasen2006} as they become very inefficient
in this regime. Instead, we rely on the radiation hydrodynamics approach
\stella to predict early observables for \snias{} in this work. This code has
been developed by \citet{Blinnikov1993}, \citet{Blinnikov1998} and
\citet{Blinnikov2000} and has been successfully used to study a variety of
aspects of supernova research.  Prominent examples include studying SN~1987A
\citep[][]{Blinnikov2000}, supernova shock breakouts
\citep[e.g.][]{Blinnikov2011}, calculating light curves for \snias{}
\citep{Blinnikov2006, Woosley2007}, exploring super-luminous supernovae
\citep{Baklanov2015, Sorokina2016a}, investigating pair-instability supernovae
\citep{Kozyreva2017} or studying ejecta--CSM interaction in the context of
super-Chandrasekhar \snias{} \citep{Noebauer2016}. 

We briefly review some of the essential characteristics of \stella, which are
of relevance for the current study. More details and in-depth descriptions of
the involved techniques may be found in \citet{Blinnikov1998, Blinnikov2006}.
\stella is a one-dimensional radiation hydrodynamics code, which employs a
fully implicit solution procedure and tackles the radiation transfer aspect of
the problem with a multi-group variable Eddington factor scheme. Hereby, the
effects of atomic line interactions, electron scattering, inverse
bremsstrahlung and photoionization are taken into account. In particular, the
contribution of roughly 160,000 lines, taken from the \citet{Kurucz1995} data
base, is included in the opacity calculation by employing the expansion opacity
formalism of \citet{Friend1986}. Local thermodynamic equilibrium (LTE) is
assumed when determining the ionization and excitation balance. However, no
such assumptions are imposed on the radiation field itself, whose
non-equilibrium evolution is followed. In conjunction with the detailed
multi-group treatment, a simple single-group diffusion scheme is also included
to track the transport and deposition of $\gamma$-radiation energy released in
the radioactive decays \citep[cf.][]{Blinnikov2006}. For most calculations
presented here, only the energy generation in the most important decay chain in
\snias{} is taken into account, namely the decay of \nuc{56}{Ni} to
\nuc{56}{Co} and \nuc{56}{Fe}. However, for studying the double-detonation
scenario (cf.\ Section~\ref{sec:subchmodels}), the additional decay channels
$\nuc{52}{Fe} \rightarrow \nuc{52}{Mn} \rightarrow \nuc{52}{Cr}$ and
$\nuc{48}{Cr} \rightarrow \nuc{48}{V} \rightarrow \nuc{48}{Ti}$ have been
implemented into \stella{}. Half-lives and the amounts of energy that are
released in form of $\gamma$-radiation and particles have been adopted from
\citet{Dessart2014a}. The latter contribution is treated as an instantaneous
heating term in \stella{}. In all calculations a grey specific absorption
cross-section of $\kappa = \SI{0.05}{cm^2.g^{-1}}$ for $\gamma$-radiation is
used in the single group diffusion scheme.

\section{Models}
\label{sec:models}

In this work, we aim at predicting and investigating early-time observables of
\snias{}, in particular light curves, for a set of explosion models.  Hereby,
we primarily focus on models that produce about $0.55 -
0.60\,\mathrm{M}_{\sun}$ of radioactive material and are thus broadly
compatible with the brightness of normal \snias{}, like SN~2011fe for example
\citep{Pereira2013}.  Based on these criteria, we selected two
Chandrasekhar-mass ($M_{\mathrm{Ch}}$) explosion models and three
sub-Chandrasekhar models as the basis for this work. In the investigation of
specific early light curve features (see Section~\ref{sec:power_law_rise}), we
also focus on the consequences of mixing in the ejecta. For this purpose, a
small suite of toy models was constructed and two pure deflagration explosion
models were considered. The latter two serve as an example of completely mixed
ejecta. With the exception of the W7 model \citep{Nomoto1984}, all original
explosion calculations have been performed by the \snia{} group formerly based
at the Max Planck Institute for Astrophysics (MPA). These models are part of
the recent public release of the
\textsc{Hesma}\footnote{\url{https://hesma.h-its.org/}} data base
\citep{Kromer2017}. In the following, all models are briefly introduced
individually.

\subsection{Explosion models for normal \snias{}}

\subsubsection{Carbon deflagration: W7}
\label{sec:W7model}

Despite its parametrized description of the thermonuclear burning process in
1D, the ejecta structure of the so-called W7 model of \citet{Nomoto1984} has
been quite successful in reproducing observations of normal \snias{}
\citep{Branch1985, Jeffery1992, Hoeflich1995, Nugent1997, Lentz2001, Salvo2001,
Baron2006, Gall2012} and is still considered a standard theoretical reference
model for the thermonuclear explosion of $M_{\mathrm{Ch}}$ CO WDs. In this
work, we use the revised version of this model published by
\citet{Iwamoto1999}, which includes an extended nucleosynthesis calculation.
The model contains $0.59\,\mathrm{M}_{\odot}$ of \nuc{56}{Ni} located in a
shell roughly between $\SI{2700}{km.s^{-1}}$ and $\SI{11000}{km.s^{-1}}$. While
the innermost regions are composed of stable iron, a region rich in
intermediate mass elements (IME) lies on top of the nickel shell. Finally, the
outermost zones of the W7 model have not burnt and are thus composed entirely
of CO fuel. 

\subsubsection{Delayed detonation: N100}
\label{sec:N100model}

Delayed detonations \citep{Blinnikov1986,Blinnikov1987,Khokhlov1991} have long
been considered as the most promising mechanism to explain normal \snias{}
\citep[e.g.][]{Roepke2012} for $M_{\mathrm{Ch}}$ CO WD progenitors. In this
scenario, the thermonuclear flame first propagates as a subsonic deflagration
which later transitions into a detonation.  \citet{Seitenzahl2013} have
explored this scenario with 3D hydrodynamical simulations, investigating a
range of different possible realizations. Following \citet{Roepke2012}, who
compared the N100 model of this series to SN~2011fe, we use this model for this
work as well. In this realization of the delayed detonation scenario, the
initial deflagration is triggered in 100 ignition spots distributed randomly in
the centre of a $M_{\mathrm{Ch}}$ CO WD. After the deflagration and successive
detonation have traversed the WD and burning has ceased, a total of
$0.60\,\mathrm{M}_{\odot}$ of radioactive nickel have been synthesized in the
explosion (cf.\ Table~\ref{tab:model_summary}).

\subsubsection{Violent merger}
\label{sec:mergermodel}

The violent merger scenario \citep[e.g.][]{Pakmor2010, Pakmor2012, Pakmor2013,
Moll2014, Tanikawa2015} constitutes an alternative route to \snias{}. In this
scenario, the merger of two sub-Chandrasekhar mass CO WDs in a close binary,
triggers a carbon detonation in the more massive of the WDs, which, in turn,
disrupts the whole system. \citet{Pakmor2012} presented a binary configuration
consisting of a $0.9$ and a $1.1\,\mathrm{M}_{\sun}$ CO WD, which produced
$0.61\,\mathrm{M}_{\sun}$ of $^{56}$Ni and explains the optical spectra of
normal \snias{} reasonably well (\citealt{Roepke2012}; note, however, that the
predicted polarization signal for violent mergers is significantly stronger
than observed in normal SNe~Ia \citealt{Bulla2016}). For this work, we adopt
this model as an example of the merger scenario.

\subsubsection{Sub-Chandrasekhar mass detonation models}
\label{sec:subchmodels}

\citet{Shigeyama1992} and \citet{Sim2010} showed that centrally ignited
detonations in sub-Chandrasekhar mass CO WDs successfully reproduce important
characteristics of normal \snias{} thus highlighting the potential significance
of sub-Chandrasekhar models as progenitors for \snias{}. However, more
realistic simulations \citep[e.g.][]{Kromer2010, Woosley2011}, where the carbon
detonation in the core of the sub-Chandrasekhar mass WD is triggered by an
initial detonation in a He surface layer that has been accreted from a binary
companion (so-called double-detonation mechanism, e.g.\ \citealt{Iben1987}),
are less promising. In these models, iron-group elements (IGE) rich ashes from
the initial detonation in the He surface layer typically lead to observational
fingerprints that are not observed in normal \snias{} (but see
\citealt{Kromer2010, Shen2014} for mechanisms to partially suppress the IGE
pollution.)

Here, we pick two models to investigate the early-time observables of both
double detonations and bare CO detonations in sub-Chandrasekhar mass WDs. For
the double detonation scenario (SubChDoubleDet), we take model 3 of
\citet{Fink2010} and \citet{Kromer2010}, which yields $0.55\,\mathrm{M}_{\sun}$
of \nuc{56}{Ni} from an initial WD with a $1.03\,\mathrm{M}_{\sun}$ CO core and
a He shell of $5.5\times\SI{e-2}{\mathrm{M}_{\sun}}$. In the initial
helium-shell detonation, $\SI{1.7e-3}{\mathrm{M}_{\sun}}$ of \nuc{56}{Ni},
$\SI{5.6e-3}{\mathrm{M}_{\sun}}$ of \nuc{52}{Fe} and
$\SI{4.0e-3}{\mathrm{M}_{\sun}}$ of \nuc{48}{Cr} have been synthesized close to
the ejecta surface. As a bare CO detonation (SubChDet), we use the model of a
detonation of a $1.06\,\mathrm{M}_{\sun}$ WD that yields
$0.56\,\mathrm{M}_{\sun}$ of \nuc{56}{Ni} (model 1.06 of \citealt{Sim2010}).

\subsection{Models with increased mixing}

\subsubsection{N100 toy model suite}
\label{sec:N100_toymodels}

\begin{figure}
  \centering
  \includegraphics{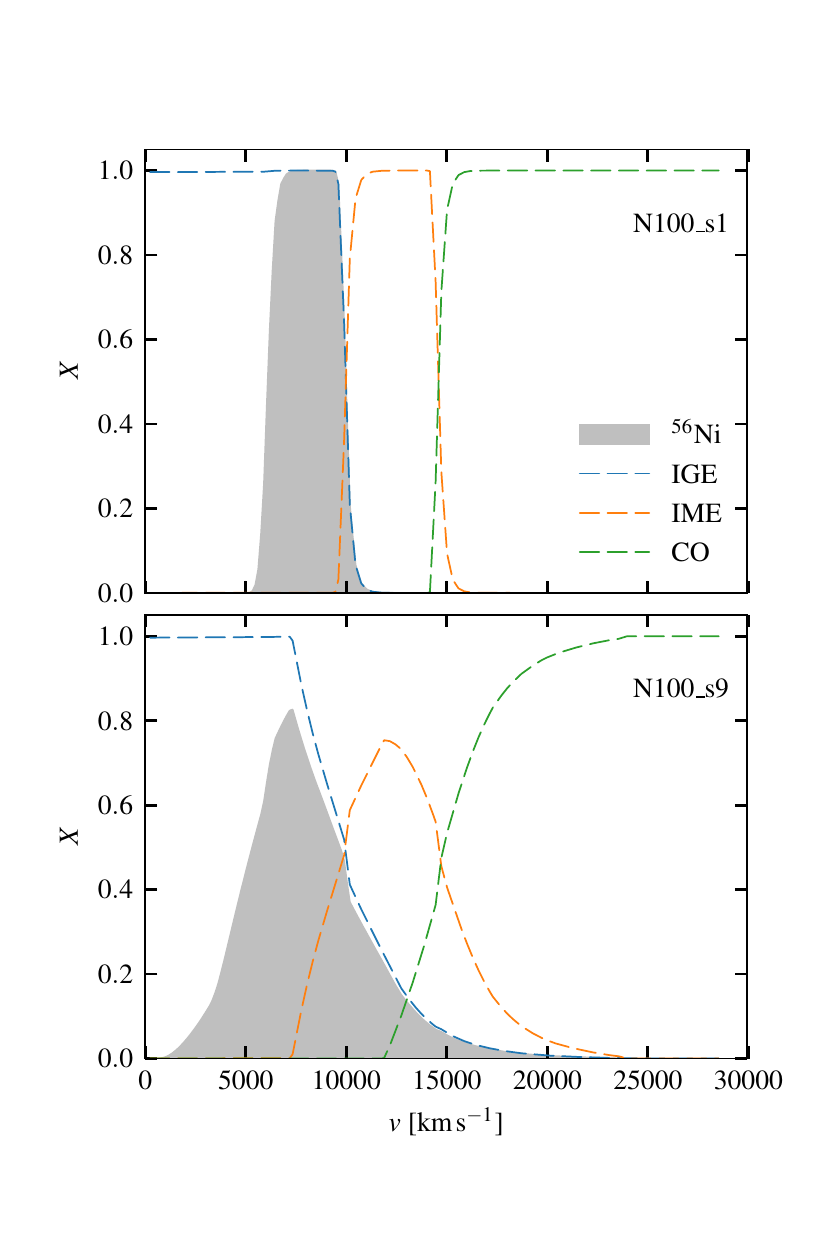}
  \caption{Composition in the toy model suite constructed for exploring the
    effects of stratification. The upper panels shows the ejecta structure in
    the $k=1$ model of the series. The composition of the model with the
    strongest smoothing ($k=9$) is illustrated in the lower panel.}
  \label{fig:N100s_comp}
\end{figure}

Mixing in the ejecta has some important consequences on the shape of the early
\snia{} light curves, as will be shown later.  To reveal the correlations
between mixing and certain features in the early light curves systematically, a
small suite of toy models is considered. This model series, which adopts a
simplified ejecta composition, has been constructed by Klauser (B.A.~Thesis,
LMU) to study the effect of mixing of iron-group elements on synthetic light
curves around maximum light and has also been used by \citet{Sasdelli2017}. For
the base model of the series, the density profile of the N100 model is adopted
(see Section~\ref{sec:N100model}). Also, the total masses of \nuc{56}{Ni},
IGE\footnote{All elements heavier than calcium.}, IME and unburned fuel (CO)
were taken from N100.  These elemental groups are then rearranged to form a
highly stratified ejecta structure. In particular, the innermost regions are
filled up entirely by stable IGEs. Adjacent to this core region, a layer
containing exclusively the entire radioactive material is located, which is
followed by an IME zone and finally the outermost regions which only contain
unburned fuel. For simplicity, it is assumed that the stable IGE layer is
composed of only of iron and that the total IME mass is distributed in equal
parts on silicon, sulphur, magnesium and calcium.  To explore the consequences
of strong layering on the light curve in more detail, variants of this model
are constructed by applying various degrees of smoothing to the base ejecta
composition. In particular, a Gaussian smoothing kernel is used, whose strength
is controlled via the standard deviation $\sigma$ of the Gaussian. In total, a
suite of five toy models is generated by using 
\begin{equation}
  \sigma = k \times \SI{335}{km.s^{-1}}
  \label{eq:smoothing_len}
\end{equation}
and $k$ in $[1, 2, 3, 6, 9]$. The impact of this smoothing on the model
composition is highlighted in Fig.~\ref{fig:N100s_comp}, which compares the
distribution of the main elemental groups in the models with $k=1$ and $k=9$.

\subsubsection{Pure deflagrations}

Subsonic, turbulent burning introduces strong mixing. Thus, we also consider
two pure deflagrations from the model suite studied by \citet{Fink2014}. In
that work, a range of possible realizations of pure deflagrations in
$M_{\mathrm{Ch}}$ CO WDs were explored in detailed three-dimensional
calculations. With the N5def and N1600Cdef models, we adopt two examples from
the ends of the range of deflagration strengths considered. The N5def model is
of particular interest since it belongs to the family of so-called failed
deflagrations where the burning is not energetic enough to unbind the entire WD
and a bound remnant remains. It has been shown that such models are good
candidates for the faint SN~2002cx-like objects \citep[e.g.][]{Kromer2013}.

\subsection{Mapping into \stella}
\label{sec:mapping_stella}

With the exception of the W7 and the double-detonation model, the original
explosion calculations have been performed until $t_{\mathrm{stop}} =
\SI{100}{s}$ after explosion.  At this time the ejecta are in almost perfect
homologous, i.e.\ force free, expansion. In particular, \citet{Roepke2005}
showed that already at $t \sim \SI{5}{s}$ deviations from homology are very
small. Thus, input models for \snias{} radiative transfer calculations are
typically constructed in perfect homologous expansion
\citep[e.g.][]{Kasen2009,Hillier2012}.  We follow this practice when mapping
into \stella. In particular, we project the different explosion models on to a
uniform spherical grid consisting of $200$ initially equidistant cells. In this
process, small deviations in the mass distribution may occur, leading to total
ejecta masses slightly different from the values reported in the original
publications of the respective explosion models. We start all \stella
calculations at $t_{\mathrm{exp}} = \SI{e4}{s}$ after explosion. We bridge the
very early phase from $t_{\mathrm{stop}}$ to $t_{\mathrm{exp}}$, when the
ejecta are very optically thick and virtually opaque to radiation, by
homologously expanding the models. For simplicity, we assume an isothermal
stratification in the ejecta with $T = \SI{2e3}{K}$ at the simulation start.
This value is somewhat arbitrary but overall in rough agreement with the
adiabatic cooling effect due to volume expansion, which is expected to have
reduced the temperature of the initially hot ejecta material, heated by
thermonuclear burning and shock heating. A total of $200$ frequency groups are
used, which are logarithmically distributed between $\SI{6.1e13}{Hz}$ and
$\SI{7.6e16}{Hz}$, corresponding to the wavelength range roughly from
$\SI{39}{\text{\AA}}$ to $\SI{49000}{\text{\AA}}$. We have explicitly checked
that increasing the spectral and spatial resolution does not change the results
significantly at the example of the W7 model.

\begin{figure}
  \centering
  \includegraphics[]{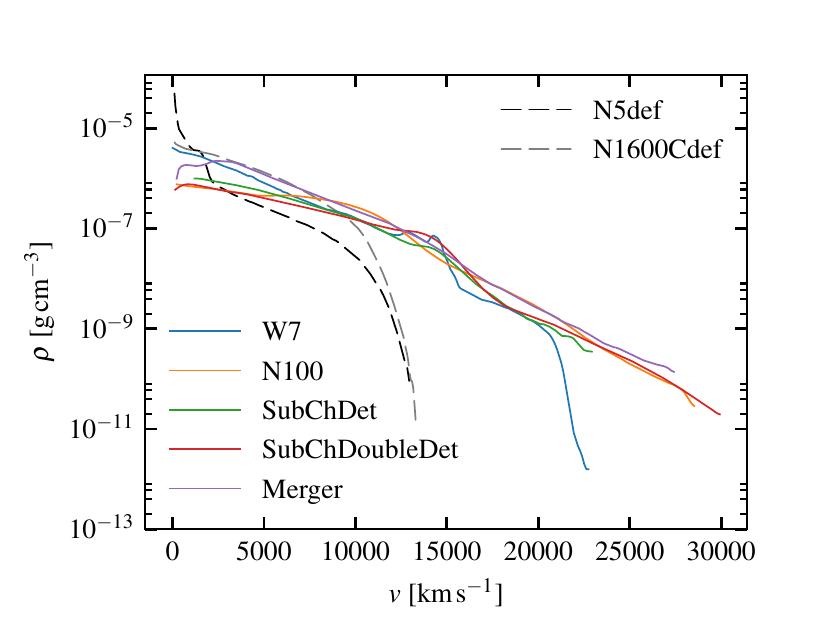}
  \caption{Overview of the main model series in terms of their density profiles
    after mapping on to a one-dimensional spherical grid (where needed) at
    $t_{\mathrm{exp}} = \SI{e4}{s}$ after explosion. For completeness, the two
    pure deflagration models (dashed lines), which are used in a later part of
  this work, are included as well.}
  \label{fig:model_dens}
\end{figure}

All explosion models are summarized in Table~\ref{tab:model_summary} in terms
of their fundamental properties, and their density profiles at $t = \SI{e4}{s}$
after explosion are shown in Fig.~\ref{fig:model_dens}. 
\begin{table}
  \centering
  \caption{Key properties of the different explosion models used in this work.
    In particular, the total ejecta mass ($M_{\mathrm{tot}}$) and the mass of
    radioactive nickel ($^{56}$Ni-mass) are given in units of solar masses and
    the total kinetic energy ($E_{\mathrm{kin}}$) in $10^{51}\,\mathrm{erg}$ is
    provided. Additionally, the time until which the original explosion
    simulation was performed, $t_{\mathrm{stop}}$, is indicated in seconds. In
    the first part, the group of models suitable for normal \snias{} are
    listed. For completeness, the properties of the two pure deflagration
    models used later on (see Section~\ref{sec:power_law_rise}) are included at
  the bottom.} 
  \label{tab:model_summary}
  \begin{tabular}{ccccc}
    \hline
Model & $M_{\mathrm{tot}}$ & $M_{\mathrm{^{56}Ni}}$ & $E_{\mathrm{kin}}$ & $t_{\mathrm{stop}}$ \\\hline
              W7 & 1.380 & 0.587 & 1.103 &  20.0\\
            N100 & 1.406 & 0.604 & 1.424 & 100.0\\
          Merger & 1.935 & 0.613 & 1.607 & 100.0\\
        SubChDet & 1.062 & 0.559 & 0.978 & 100.0\\
  SubChDoubleDet & 1.095 & 0.549 & 1.230 &   7.8\\
           N5def & 0.353 & 0.152 & 0.131 & 100.0\\
       N1600Cdef & 1.400 & 0.320 & 0.562 & 100.0\\
    \hline
  \end{tabular}
\end{table}
The toy model suite does not appear in these summaries since their fundamental
properties and density profiles are identical to N100.

\section{Results}
\label{sec:results}

After mapping the explosion models into \stella, their radiation hydrodynamical
evolution is followed. As expected from previous studies
\citep[see][]{Pinto2000, Woosley2007, Noebauer2012}, the dynamical impact of
the radiation generated by the radioactive decay is small and the ejecta remain
almost in perfect homologous expansion. During the first $\SI{10}{d}$ after
explosion, deviations from homology remain smaller than 5 per cent in the
ejecta velocity and 10-20 per cent in the density, barring numerical effects at
the computational boundaries.

As the ejecta evolution is followed, \stella calculates in each time-step the
SED of the emergent radiation field.  Although this SED is too coarse for
spectral synthesis purposes, it is well suited for determining colour curves.
Throughout this work, light curves will be shown in terms of absolute
AB-magnitudes, which have been obtained by convolving the synthetic SED with
the respective filter transmission curves $S_x(\lambda)$
\citep[cf.][]{Bessell2012}
\begin{equation}
  M_{\mathrm{AB}} = -2.5 \log \frac{\int \mathrm{d}\lambda
  F_{\lambda}(\lambda) S_{x}(\lambda) \lambda}{\int \mathrm{d} \lambda S_{x}(\lambda) c /\lambda}
  \label{eq:ab_magnitudes}
\end{equation}
Here, the synthetic flux $F_{\lambda}(\lambda)$ at a distance of $\SI{10}{pc}$
in units of $[\mathrm{erg\,s^{-1}\,cm^{-2}\,\text{\AA}^{-1}}]$ enters.

\begin{figure*}
  \centering
  \includegraphics{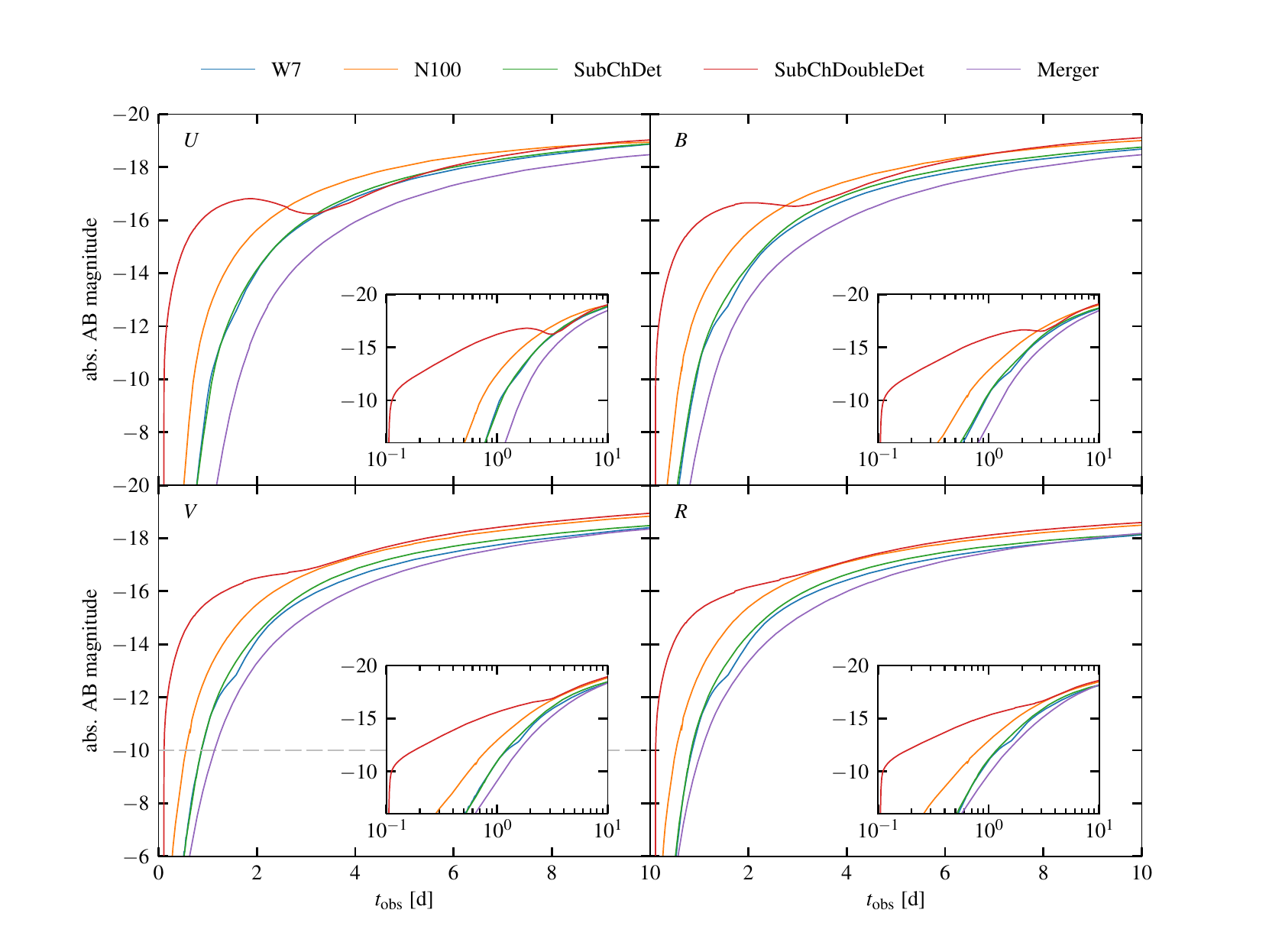}
  \caption{Overview of the synthetic light curves for the different models in
    the Bessell $U$ (upper left-hand panel), $B$ (upper right-hand panel), $V$
    (lower left-hand panel) and $R$ (lower right-hand panel) passbands during
    the first $\SI{10}{d}$ after explosion. The grey dashed horizontal line
    marks the limiting $V$ band magnitude, which is used in the discussion of
    the different rise behaviour in Section~\ref{sec:rise}. Insets in each
    panel show the respective light curves on a logarithmic time~scale. The
    steep rise at $0.12$\,d in model SubChDoubleDet is an artefact resulting
    from our choice of $t_{\mathrm{exp}} = \SI{e4}{s}$ (see discussion in
    Section~\ref{sec:limitations}).} 
  \label{fig:earlyrise_summary}
\end{figure*}

Fig.~\ref{fig:earlyrise_summary} shows the resulting light curves in the
Bessell $U$, $B$, $V$ and $R$ bands \citep{Bessell2012} of all models for the
first $\SI{10}{d}$ after explosion.
Overall, the light curve shape is largely similar, with the exception of the
double-detonation model. Here, the radioactive material close to the surface,
which has been synthesized in the helium-shell detonation (\nuc{48}{Cr},
\nuc{52}{Fe}, \nuc{56}{Ni}) leads to prominent first peaks or shoulders in the
light curves in all investigated bands. Although most models show similarly
shaped early light curves, differences in the rise are clearly seen. If we
consider the time when the light curves reach an absolute magnitude of
$M_{\mathrm{AB}} = -10$ in the $V$ band, a spread of about $\Delta t \sim
\SI{0.6}{d}$ is observed. 

Contrary to the commonly used assumption \citep[e.g.][]{Riess1999, Conley2006,
Strovink2007, Hayden2010, Nugent2011, Firth2015}, the light curves do not
follow a strict power law (not even a broken one) during these early phases.
This becomes evident when displaying the light curve versus a logarithmic time
axis, as done in insets shown in Fig.~\ref{fig:earlyrise_summary}. A power-law
evolution would follow a straight line in this visualization, which is not the
case for all models included in this comparison during the first $\SI{10}{d}$.
This has important implications for reconstructing the time of explosion from
observed early photometric data (see Section~\ref{sec:exp_time})

For the W7 model a small but visible break in the light-curve evolution around
$t_{\mathrm{obs}} = \SI{1}{d}$ is observed in all but the $U$ band. We explore
the origin of this light curve break in detail in
Section~\ref{sec:stratification}.

\section{Discussion}
\label{sec:discussion}

\subsection{Rising phase}
\label{sec:rise}

\begin{figure}
  \centering
  \includegraphics{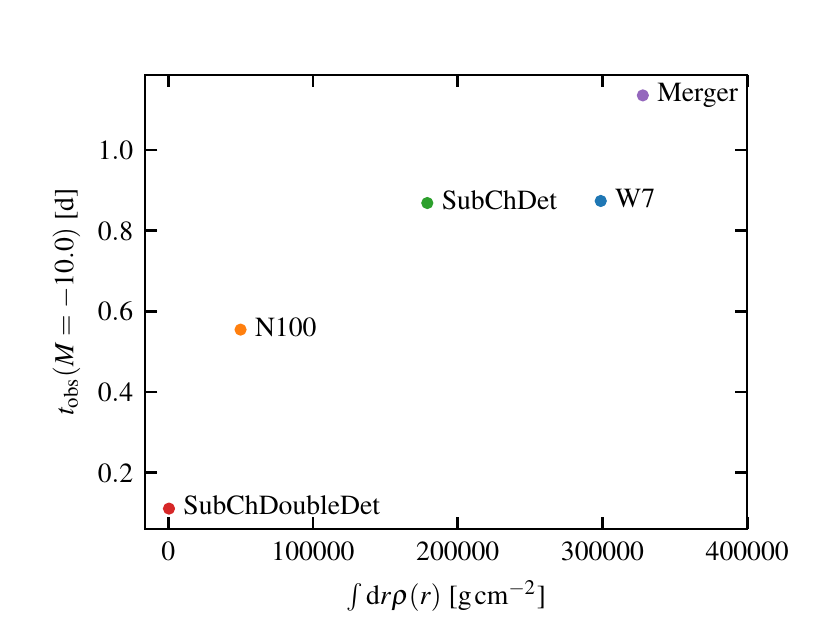}
  \caption{Correlation between the time since explosion until an absolute
    magnitude of $M=-10$ is reached in the $V$ band and the column density from
    the ejecta surface down to the \nuc{56}{Ni} zone. In particular,
    equation~\ref{eq:column_density} is used to calculate this model property
    at $t = \SI{e4}{s}$ after explosion.}
  \label{fig:trise_colrho}
\end{figure}

While all models that do not contain radioactive material close to the ejecta
surface (i.e.\ with the exception of the double-detonation model) exhibit
similarly shaped light curves, a different rise behaviour is observed. In
particular, the various model light curves reach certain limiting absolute
magnitudes at different times after explosion. The largest difference exists
between the N100 and the merger model, which surpass the (arbitrarily chosen)
limiting magnitude of $M_{\mathrm{AB}} = -10$ in the $V$ band roughly $\Delta t
= \SI{0.6}{d}$ apart. The remaining models fall between these two extremes.
The reason for this different behaviour in the light curve rise lies in varying
ejecta masses and different \nuc{56}{Ni} distributions. These properties can be
combined into the column density 
\begin{equation}
  n = \int_{r_{\mathrm{Ni}}}^{r_{\mathrm{out}}} \mathrm{d}r \rho(r)
  \label{eq:column_density}
\end{equation}
measured from the ejecta surface at $r_{\mathrm{out}}$ and to the outermost
ejecta zones where \nuc{56}{Ni} is still abundant (we selected
$X_{\nuc{56}{Ni}} \ge 0.01$ here). As shown in Fig.~\ref{fig:trise_colrho}, a
clear correlation exists between this model property (measured at $t =
\SI{e4}{s}$ after explosion)  and the time at which the limiting magnitude is
reached.
Hereby, models with less material on top of the \nuc{56}{Ni}-rich regime rise
earlier, since the radiation generated in the radioactive decay can more easily
diffuse out of the ejecta. 

\subsection{Surface Radioactivity}
\label{sec:surface_radioactivity}

\begin{figure}
  \centering
  \includegraphics{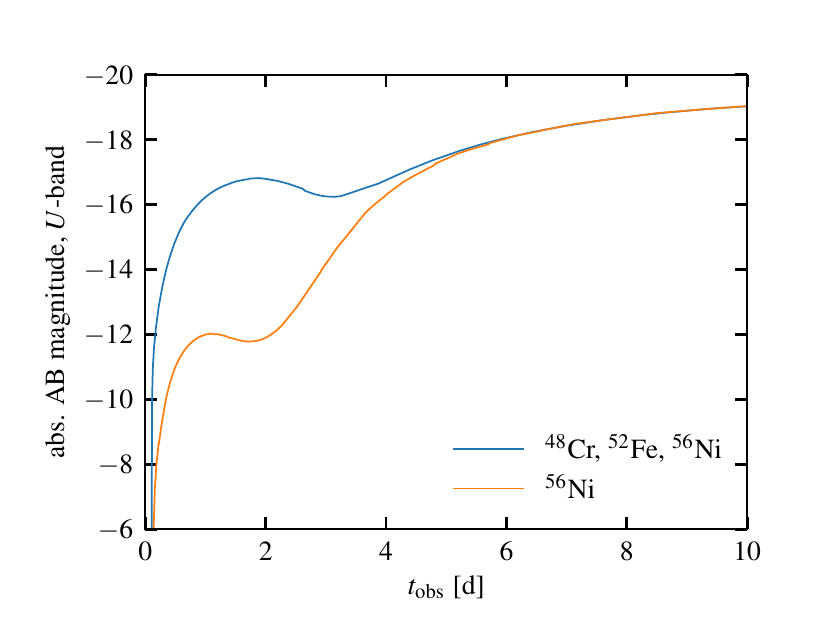}
  \caption{Comparison between $U$ band early light curve for the full
    SubChDoubleDet model and a version which only accounts for the energy
    generation in the \nuc{56}{Ni} decay chain. Note that the peak of the
    surface radioactivity peak happens later if the additional decay chains are
    included. This is due to the decay of the \nuc{48}{V}, which has a
    half-life of $t_{1/2} = \SI{15.973}{d}$ and releases a fair amount of
  energy.}
  \label{fig:model3_wwo_ni}
\end{figure}

As seen in Fig.~\ref{fig:earlyrise_summary}, even small amounts of radioactive
material close to the surface of the SN ejecta lead to a very early rise of the
light curves. For the double-detonation model investigated here,
$\SI{1.13e-2}{\mathrm{M}_{\sun}}$ of radioactive material has been synthesized
in the He-shell detonation (see Section~\ref{sec:subchmodels}). Since this
material is clearly separated from the \nuc{56}{Ni} in the central ejecta
regions, the corresponding radioactive decays lead to a first peak in the $U$
and $B$-band light curves and a pronounced shoulder in $V$ and $R$. In the
model presented here, the prompt surface radioactivity signal is dominated by
the decays of \nuc{52}{Fe} and \nuc{48}{Cr} that occur much faster than the
\nuc{56}{Ni} chain. However, qualitatively similar results are also obtained
with only \nuc{56}{Ni} close to the surface and separated from the core
material as shown in Fig.~\ref{fig:model3_wwo_ni}.
Here, the full SubChDoubleDet model is compared with a modified version, in
which only the energy generation due to the \nuc{56}{Ni} decay chain is taken
into account. As shown in the figure, already the \nuc{56}{Ni} produced in the
He-shell detonation produces a prominent early peak in the $U$-band light curve
\citep[see also][who found a similar behaviour for UV light curves of a
double-detonation model]{Blinnikov2000a}.

\begin{figure}
  \centering
  \includegraphics{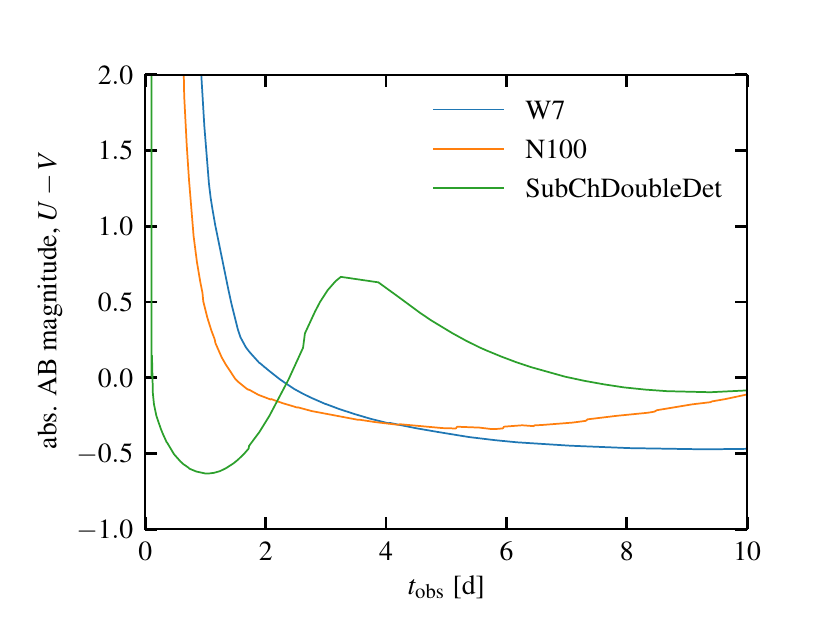}
  \caption{$U-V$ colour evolution in the double detonation (SubChDoubleDet),
  the N100 and the W7 models. The prompt emission due to the radioactive
material at the ejecta surface leads to a blue colour in the double-detonation
model very early on.}
  \label{fig:earlyrise_colours_UV}
\end{figure}

Surface radioactivity has also a profound influence on the early colour
evolution as demonstrated in Fig.~\ref{fig:earlyrise_colours_UV} in terms of
$U-V$.
During the first $\sim \SI{2}{d}$ in which the light output is powered by the
decay of radioactive material close to the surface, the supernova appears very
blue. As the prompt emission fades, the supernova becomes red again until the
colour evolution is dominated by the radiation emanating from energy generation
processes in the inner ejecta regions.

Qualitatively, the early evolution of the supernova colour and of the light
curves in the different bands due to surface radioactivity is very similar to
the predictions by \citet{Kasen2010} concerning the signatures induced by the
interaction of the ejecta with a companion. Also, interaction between ejecta
and CSM may lead to similar early peaks and shoulders, as demonstrated by
\citet{Piro2016}. Consequently, we caution that not every blue excess detected
very early on has to be interpreted as an indication for interaction between
ejecta and CSM or a companion, but may also point towards radioactive material
close to the surface. Information about the early SED of the exploding object
may help to break this degeneracy. 

\subsection{Stratification}
\label{sec:stratification}

A peculiar result of the early light curve comparison in
Section~\ref{sec:results} is certainly the small but prominent break that the
W7 model exhibits (cf.\ Fig.~\ref{fig:earlyrise_summary}). This kink in the
light curve cannot be caused by surface radioactivity since the entire
$^{56}$Ni is confined to regions with $v \lesssim \SI{1.1e4}{km.s^{-1}}$ and
thus located quite far from the ejecta surface. Furthermore, we can exclude a
link between this feature and the steep density drop-off that W7 exhibits in
the outer ejecta regions and which is not observed in other \snia{} models
(cf.\ Fig.~\ref{fig:model_dens}). For this purpose, we constructed a W7-like
toy model in which the density over the entire velocity range of the original
model is replaced by the exponential profile
\begin{equation}
  \rho(v) = \rho_0 \exp(-b v).
  \label{eq:w7_toy_exp_density}
\end{equation}
The composition of this toy model is adopted from the original W7 model. In
this procedure, the parameters $\rho_0 = \SI{5.44e-6}{g.cm^{-3}}$ and $b =
\SI{3.67e-4}{s.km^{-1}}$ are chosen such that the total mass remains constant.
The masses of the various elements merely change slightly since the density
differences are only significant in the outermost ejecta regions. The light
curves calculated for this toy model still exhibit an almost identical early
light curve break as the original W7 model.

\begin{figure}
  \centering
  \includegraphics{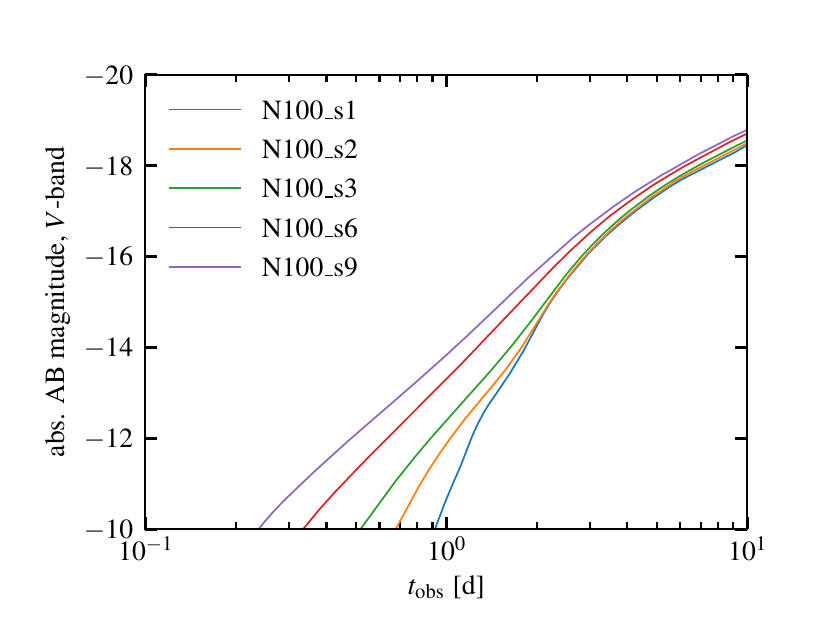}
  \caption{Early time $V$-band light curve for the suite for toy models derived
    from the N100 model to study the effect of stratification. While a W7-like
    early break is seen for the models with strong layering of the different
    elemental groups, it disappears with stronger mixing. }
  \label{fig:earlyrise_V_N100s_cmp}
\end{figure}

Instead, this light curve feature seems to be associated with the strong
stratification and layered composition in the W7 model. To explore this
correlation, the toy model suite based on the N100 model in which the degree of
internal mixing is gradually increased is considered (see
Section~\ref{sec:N100_toymodels}). In Fig.~\ref{fig:earlyrise_V_N100s_cmp}, the
resulting $V$-band light curves for these models as calculated with \stella{}
are shown. 
While the highly stratified model strongly exhibits a light curve kink as seen
in W7, the kink gradually disappears as the degree of mixing increases. 

\begin{figure}
  \centering
  \includegraphics{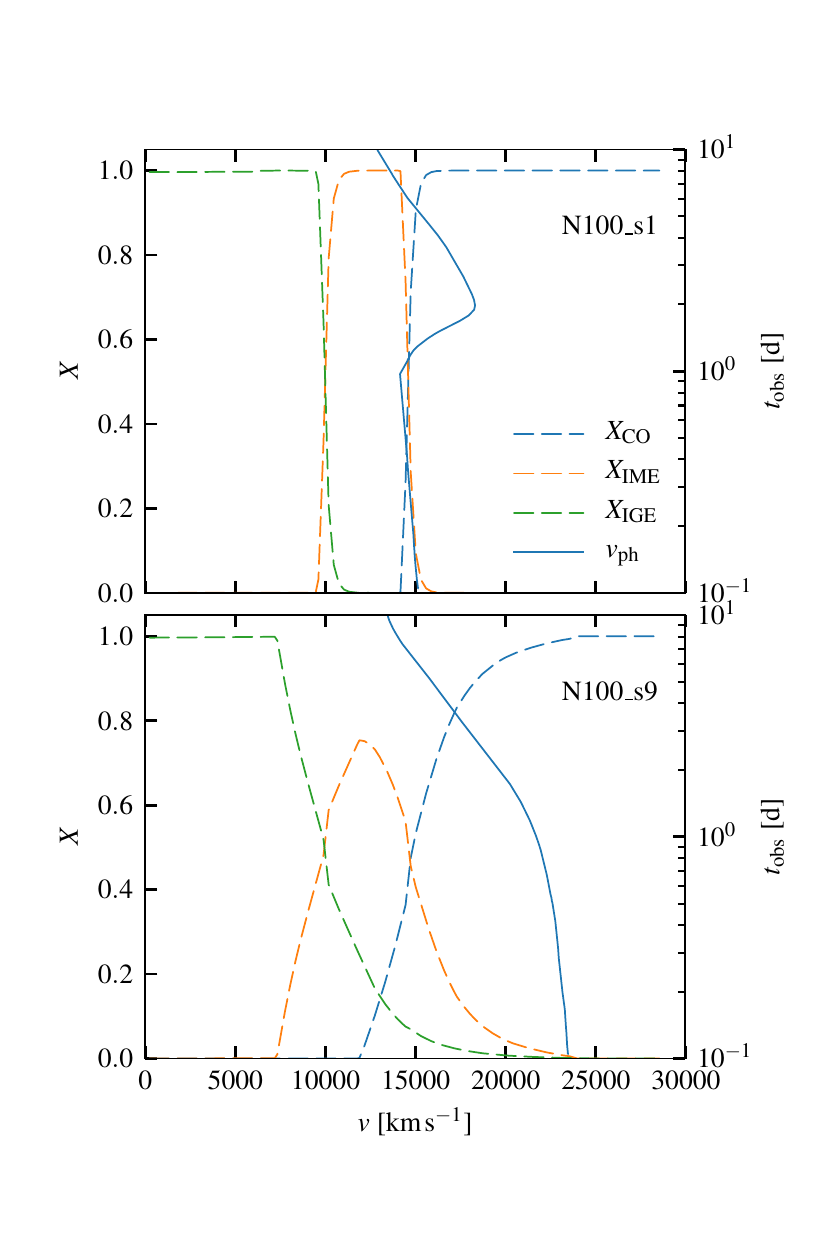}
  \caption{Illustration of the location of the photosphere ($\tau = 2/3$
    surface, blue solid line in the N100\_s1 model (top panel). Its time
    evolution is tracked on the right-hand-side y-axis from the bottom to top,
    while the abundances of the different elemental groups (dashed lines) are
    shown on the left-hand-side y-axis. During very early phases, the
    photosphere coincides with the IME--CO interface. Only after the radiation
    which is trapped inside this region can escape, the photosphere first
    expands and follows its normal receding evolution (see also
    Fig.~\ref{fig:phot_evolution}). In the bottom panel, the situation is shown
    for the N100\_s9 model. Since there is a smooth transition between IME and
    CO layers, heating can already occur in the outer regions early on and the
    photosphere is located close to the ejecta surface and monotonously recedes
    from the beginning of the \stella{} calculation.}
    \label{fig:N100s_phot_comp_evol}
\end{figure}

\begin{figure}
  \centering
  \includegraphics{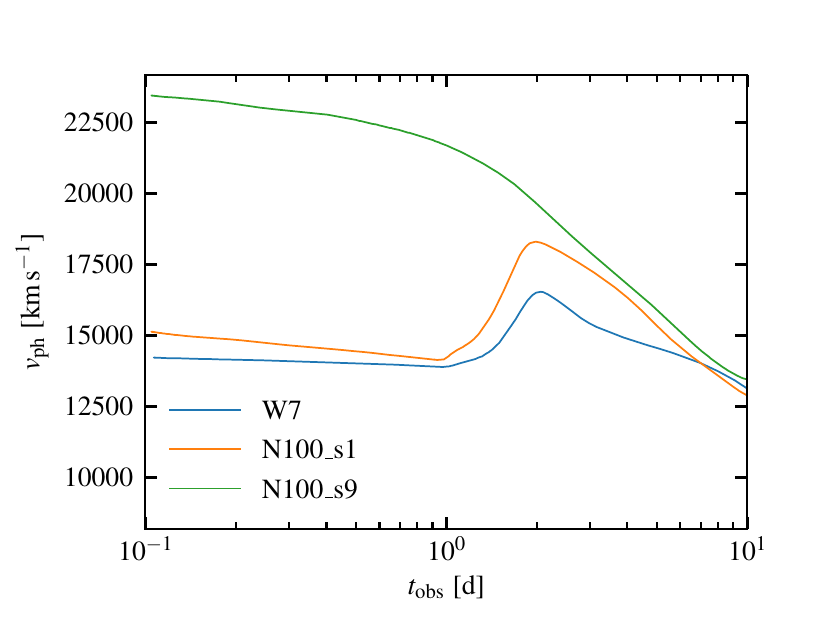}
  \caption{Comparison of the evolution of the photosphere ($\tau=2/3$) for a
    strongly stratified and a well-mixed model of the toy model series. In
    addition, the situation in W7 is included. The light curve break observed
    in W7 and strongly stratified models, in general, coincides with the phase
  during which the photosphere expands away from the IME/CNO interface and then
turns into the normal receding evolution.}
  \label{fig:phot_evolution}
\end{figure}

Physically, the light curve break is associated with the strong variations in
opacity at sharp interfaces between IME and CO regions. In strongly stratified
models, radiation generated in the ejecta core due to radioactive decays is
trapped and can only slowly leak out of the IME layer.  During these early
phases the `photosphere' is located at the IME/CO interface and remains there
as illustrated in Fig.~\ref{fig:N100s_phot_comp_evol}.
Here, and in the following the photosphere is defined as the location at with
the optical depth for radiation at the centre of the $B$-passband reaches
$\tau=2/3$.  Only after the ejecta have expanded to make the IME layers
sufficiently transparent,  radiation can break out and heat the outer regions.
This process temporarily shifts the photosphere outwards.  However, shortly
after this `break-out' the photosphere follows its expected evolution and
gradually recedes as the ejecta expand and cool. This non-monotonous evolution
of the photosphere does not occur in well-mixed ejecta since the opacity drops
off more regularly in such a configuration and radiation can continuously leak
out from the centre (see Fig.~\ref{fig:N100s_phot_comp_evol}). These
differences in the photosphere evolution are again summarized in
Fig.~\ref{fig:phot_evolution}.

\subsection{Power-law rise}
\label{sec:power_law_rise}

\begin{figure}
  \centering
  \includegraphics{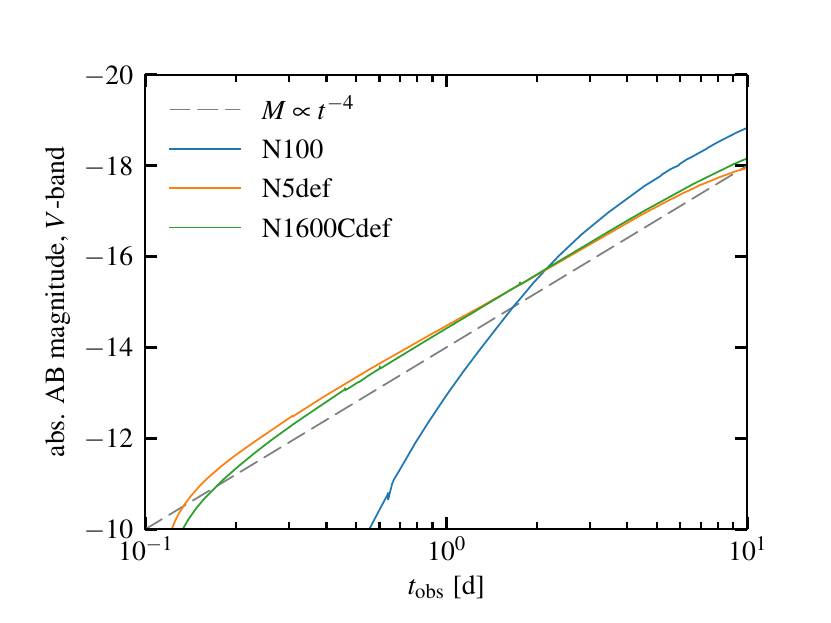}
  \caption{Early light curve for the two deflagration models N5def and
    N1600Cdef in comparison. The logarithmic display clearly highlights that
    the pure deflagration models follow a power-law rise, with an exponent of
    $-4$ in terms of absolute magnitudes. Such a power law rise is shown as the
    dashed grey line for visual comparison.  This translates into a luminosity
    evolution of $L \propto t^{1.6}$. For reference, the corresponding results
    for the N100 model are included as well.}
  \label{fig:deflagration_early_lc}
\end{figure}

None of the models for standard \snias{} investigated in
Section~\ref{sec:results} produce early light curves that follow a strict power
law.  However, the calculations with the suite of toy models presented in
Section~\ref{sec:stratification}, demonstrate that the light curves approach a
power-law rise as the degree of mixing in the ejecta increases (see
Fig.~\ref{fig:earlyrise_V_N100s_cmp}). We explore whether this trend
generalizes to pure deflagration models, since this subsonic turbulent burning
mode typically induces strong mixing in the ejecta. For this purpose, we
calculate early light curves for one-dimensional representations of the pure
deflagration models N5def and N1600Cdef, which represent extreme ends of the
range of ignition configurations investigated by \citet{Fink2014}. As
demonstrated in Fig.~\ref{fig:deflagration_early_lc}, the strong mixing in
these models indeed leads to a nearly perfect power-law rise of the light curve
during the first ten days. 
Specifically, both models roughly follow an $L \propto t^{1.6}$ rise in the $V$
band.  The same behaviour holds also in other bands, particularly in the $U$,
$B$ and $R$ bands.  

Although the pure deflagration scenario is not a suitable candidate for
explaining normal \snias{}, mainly due to its difficulty in producing large
amounts of radioactive material, this model is currently heavily discussed in
the context of \sniasx{} \citep[e.g.][]{Phillips2007,Jordan2012,Foley2013}. In
particular, weak realizations of the pure deflagration model, which leave a
bound remnant (`failed deflagrations' such as N5def) match observations of
these objects very well \citep{Kromer2013,Kromer2015}. Consequently, we predict
that \sniasx{} should be characterized by an early time strict power-law rise,
provided that the association with the pure/failed deflagration scenario
persists.

\subsection{Estimating explosion times}
\label{sec:exp_time}

From an observational point of view, the time of explosion of an \snia{} is
typically determined by fitting a power law to early photometric data and
extrapolating to $L=0$ \citep[e.g.][]{Riess1999, Conley2006, Strovink2007,
Hayden2010}. This common practice should be questioned given that none of our
normal \snia{} models produce early light curves, which follow a power-law
behaviour. We explore and illustrate the consequences and uncertainties of such
an explosion time determination by applying it to the synthetic \stella light
curves and confronting the reconstructed with the real time of explosion.

\begin{figure*}
  \centering
  \includegraphics{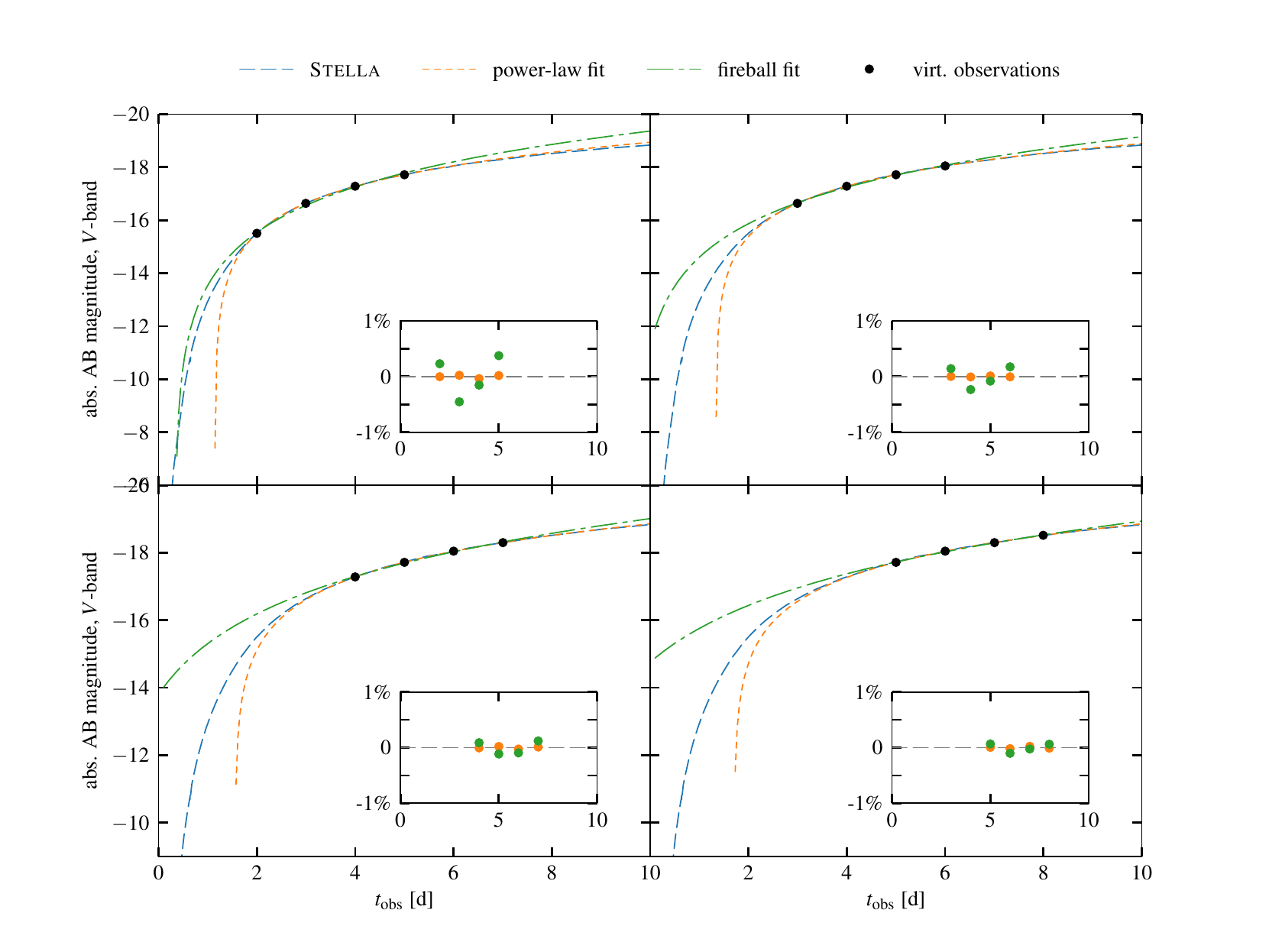}
  \caption{Illustration of fitting early virtual observations for an \snia{}
  which evolves as the N100 model. The dashed blue lines show the synthetic
  $V$-band light curve calculated with \stella. This light curve is then
  sampled to produce the virtual observations. The various panels show
  different assumptions about the time of the first observation. From left to
  right, top to bottom, $t_{\mathrm{s}}$ increased from $2$ to $\SI{5}{d}$ in
  $\SI{1}{d}$ increments. In all cases, it was assumed that after the first
  observation, additional three data points are available in $\SI{1}{d}$
  intervals. The resulting power-law and fireball model fits to these virtual
  observations are shown with dashed orange and green lines respectively. Small
  insets show the accuracy of the fit in terms of the relative deviation
  between virtual observation and fitted power-law/fireball model. In all
  cases, the deviations are below 1 per cent even though the N100 model does
  not follow a strict power law. This goodness of fit is bought by a
  reconstructed explosion time which is at times quite different from the real
  one (i.e.\ $\Delta t \neq 0$, see also Tables~\ref{tab:fitpl_params} and
  \ref{tab:fitfb_params}).}
  \label{fig:exptime_fit}
\end{figure*}

For this purpose, we extract `virtual' observations from the synthetic light
curves. Hereby, we assume that the first observation is obtained at time
$t_{\mathrm{s}}$ after explosion and that further observations are available in
daily intervals. The first four of these thus obtained epochs will be used in
the fitting process, which is similar to the situation of estimating the
explosion date for SN~2011fe \citep{Nugent2011}. In the fitting procedure, we
attempt to model these virtual observations first by a generic power law, with
an arbitrary but constant exponent (i.e.\ $L \propto t^p$)
\begin{equation}
  M_{\mathrm{PL}} = a - 2.5 p \log(t + \Delta t),
  \label{eq:powerlaw_fit_formula}
\end{equation}
and secondly by the often used fireball model \citep[see e.g.][]{Nugent2011},
which assumes that the luminosity evolves proportionally to $t^2$ and thus
\begin{equation}
  M_{\mathrm{FB}} = a - 5 \log(t + \Delta t).
  \label{eq:fireball_fit_formula}
\end{equation}
In these two expressions, $t$ and $\Delta t$ are measured in days and $a$,
$\Delta t$, and in the power-law case, $p$ are free parameters which are
marginalized. Once the fitting procedure is complete, $\Delta t$ describes any
offset between the reconstructed and the real explosion time. All model
$V$-band light curves were fitted with this procedure for $t_{\mathrm{s}}$ in
$[2, 3, 4, 5\,\mathrm{d}]$ using the differential evolution module of SciPy
\citep[version 0.19.0,][]{Jones2001}, which is based on the algorithm by
\citet{Storn1997}. Hereby, we restricted the parameter space $(a, p, \Delta t)$
to $[-40, -5] \times [0.1, 5] \times [-5, 0.99 \,t_{\mathrm{s}}]$ and minimized 
\begin{equation}
  \chi^2 = \sum_{i=1}^4 \left(M^{\mathrm{obs}}_i - M^{\mathrm{fit}}_i\right)^2.
  \label{eq:chi2}
\end{equation}
The obtained results are listed in Tables~\ref{tab:fitpl_params} and
\ref{tab:fitfb_params} and illustrated at the example of the N100 model in
Fig.~\ref{fig:exptime_fit}.
\begin{table}
  \centering
  \caption{Overview of the parameters found in the power-law fit (see
    equation~\ref{eq:powerlaw_fit_formula}) to the early light curves. In
    addition to the values of the fitted parameters, the goodness of fit in
    terms of the $\chi^2$ (see equation~\ref{eq:chi2}) is included.}
  \label{tab:fitpl_params}
  \begin{tabular}{rccccc}
  \hline
    Model & $t_{\mathrm{s}}$ (d) & $\chi^2$ & $a$ & $p$ & $\Delta t$ (d) \\\hline
    W7& $2.0$ & $\SI{8.17e-06}{}$ & $-14.92$ & $1.56$ & $1.36$ \\
    W7& $3.0$ & $\SI{4.80e-05}{}$ & $-15.17$ & $1.44$ & $1.55$ \\
    W7& $4.0$ & $\SI{4.84e-05}{}$ & $-15.80$ & $1.16$ & $2.16$ \\
    W7& $5.0$ & $\SI{2.07e-05}{}$ & $-16.01$ & $1.08$ & $2.44$ \\
    N100& $2.0$ & $\SI{6.64e-05}{}$ & $-15.73$ & $1.35$ & $1.14$ \\
    N100& $3.0$ & $\SI{6.14e-06}{}$ & $-15.95$ & $1.26$ & $1.34$ \\
    N100& $4.0$ & $\SI{3.77e-05}{}$ & $-16.15$ & $1.17$ & $1.56$ \\
    N100& $5.0$ & $\SI{3.36e-05}{}$ & $-16.25$ & $1.14$ & $1.71$ \\
    N1600Cdef& $2.0$ & $\SI{5.47e-06}{}$ & $-14.71$ & $1.45$ & $0.23$ \\
    N1600Cdef& $3.0$ & $\SI{3.21e-06}{}$ & $-14.98$ & $1.33$ & $0.51$ \\
    N1600Cdef& $4.0$ & $\SI{3.33e-06}{}$ & $-15.34$ & $1.20$ & $0.94$ \\
    N1600Cdef& $5.0$ & $\SI{1.10e-06}{}$ & $-15.71$ & $1.06$ & $1.50$ \\
    N5def& $2.0$ & $\SI{1.84e-05}{}$ & $-14.78$ & $1.37$ & $0.28$ \\
    N5def& $3.0$ & $\SI{1.53e-05}{}$ & $-15.26$ & $1.17$ & $0.79$ \\
    N5def& $4.0$ & $\SI{1.68e-06}{}$ & $-15.51$ & $1.08$ & $1.13$ \\
    N5def& $5.0$ & $\SI{1.25e-04}{}$ & $-16.34$ & $0.75$ & $2.49$ \\
    SubChDet& $2.0$ & $\SI{1.25e-03}{}$ & $-15.11$ & $1.60$ & $1.34$ \\
    SubChDet& $3.0$ & $\SI{3.13e-05}{}$ & $-15.99$ & $1.14$ & $2.01$ \\
    SubChDet& $4.0$ & $\SI{2.31e-06}{}$ & $-16.26$ & $1.01$ & $2.31$ \\
    SubChDet& $5.0$ & $\SI{1.37e-07}{}$ & $-16.39$ & $0.96$ & $2.49$ \\
    Merger& $2.0$ & $\SI{2.79e-04}{}$ & $-13.44$ & $2.25$ & $1.06$ \\
    Merger& $3.0$ & $\SI{1.39e-04}{}$ & $-14.32$ & $1.83$ & $1.56$ \\
    Merger& $4.0$ & $\SI{4.72e-05}{}$ & $-15.17$ & $1.45$ & $2.21$ \\
    Merger& $5.0$ & $\SI{1.52e-08}{}$ & $-15.61$ & $1.27$ & $2.66$ \\
  \hline
  \end{tabular}
\end{table}
\begin{table}
  \centering
  \caption{Same as Table~\ref{tab:fitpl_params} but for the fireball fit (see
  equation~\ref{eq:fireball_fit_formula}).}
  \label{tab:fitfb_params}
  \begin{tabular}{rcccc}
  \hline
    Model & $t_{\mathrm{s}}$ (d) & $\chi^2$ & $a$ & $\Delta t$ (d) \\\hline
    W7& $2.0$ & $\SI{1.14e-02}{}$ & $-14.16$ &$0.99$ \\
    W7& $3.0$ & $\SI{3.48e-03}{}$ & $-13.87$ &$0.59$ \\
    W7& $4.0$ & $\SI{2.89e-03}{}$ & $-13.60$ &$0.03$ \\
    W7& $5.0$ & $\SI{1.29e-03}{}$ & $-13.32$ &$-0.77$ \\
    N100& $2.0$ & $\SI{1.21e-02}{}$ & $-14.44$ &$0.34$ \\
    N100& $3.0$ & $\SI{3.39e-03}{}$ & $-14.09$ &$-0.27$ \\
    N100& $4.0$ & $\SI{1.41e-03}{}$ & $-13.81$ &$-1.00$ \\
    N100& $5.0$ & $\SI{6.21e-04}{}$ & $-13.59$ &$-1.71$ \\
    N1600Cdef& $2.0$ & $\SI{2.11e-03}{}$ & $-13.35$ &$-0.85$ \\
    N1600Cdef& $3.0$ & $\SI{1.08e-03}{}$ & $-13.14$ &$-1.32$ \\
    N1600Cdef& $4.0$ & $\SI{7.13e-04}{}$ & $-12.94$ &$-1.91$ \\
    N1600Cdef& $5.0$ & $\SI{5.19e-04}{}$ & $-12.74$ &$-2.70$ \\
    N5def& $2.0$ & $\SI{2.62e-03}{}$ & $-13.24$ &$-0.98$ \\
    N5def& $3.0$ & $\SI{1.80e-03}{}$ & $-12.99$ &$-1.57$ \\
    N5def& $4.0$ & $\SI{9.08e-04}{}$ & $-12.74$ &$-2.37$ \\
    N5def& $5.0$ & $\SI{1.36e-03}{}$ & $-12.42$ &$-3.67$ \\
    SubChDet& $2.0$ & $\SI{1.09e-02}{}$ & $-14.41$ &$0.99$ \\
    SubChDet& $3.0$ & $\SI{1.18e-02}{}$ & $-14.08$ &$0.56$ \\
    SubChDet& $4.0$ & $\SI{3.71e-03}{}$ & $-13.65$ &$-0.40$ \\
    SubChDet& $5.0$ & $\SI{1.34e-03}{}$ & $-13.30$ &$-1.51$ \\
    Merger& $2.0$ & $\SI{3.85e-03}{}$ & $-13.88$ &$1.24$ \\
    Merger& $3.0$ & $\SI{6.47e-04}{}$ & $-13.95$ &$1.33$ \\
    Merger& $4.0$ & $\SI{2.10e-03}{}$ & $-13.83$ &$1.14$ \\
    Merger& $5.0$ & $\SI{1.37e-03}{}$ & $-13.62$ &$0.68$ \\
  \hline
  \end{tabular}
\end{table}

\begin{figure}
  \centering
  \includegraphics{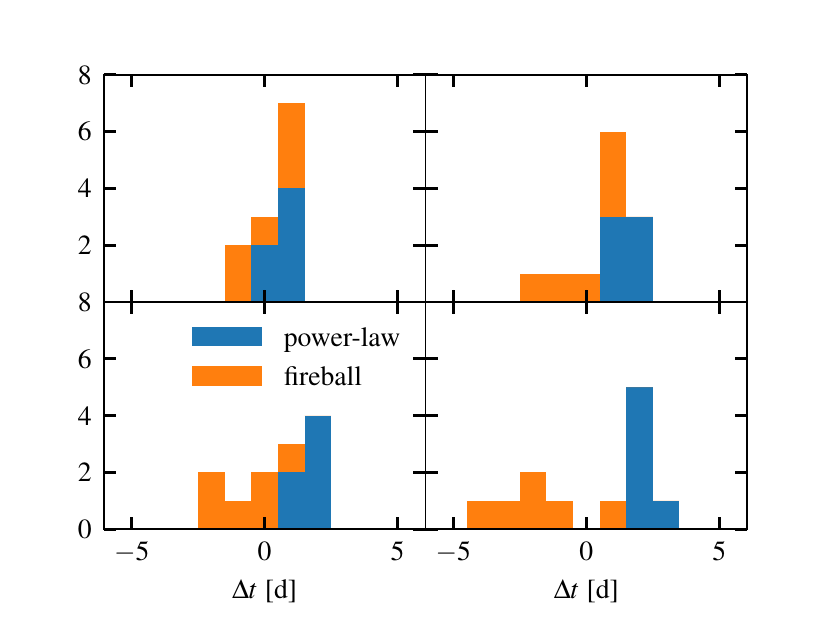}
  \caption{Illustration of the offset between reconstructed and real times of
    explosion, $\Delta t$, listed in Tables~\ref{tab:fitpl_params} and
    \ref{tab:fitfb_params}. The different panels show histograms of obtained
    $\Delta t$ for all models for different assumptions about the time of first
    observation after explosion, $t_{\mathrm{s}}$. From left to right, top to
    bottom, $t_{\mathrm{s}}$ increases from $2$ to $\SI{5}{d}$ in $\SI{1}{d}$
    increments.  The results for the power-law fit are shown in blue and the
  corresponding results for the fireball model fit in orange. This overview
clearly highlights the difficulty in reconstructing the explosion time from
fits to the early light curve.}
  \label{fig:deltat_summary}
\end{figure}

The illustration clearly demonstrates that very good fits can be achieved with
both the power-law and the fireball model despite the fact that the underlying
synthetic light curve does not follow a power-law trend. The accuracy of the
fit is illustrated in the inset in Fig.~\ref{fig:exptime_fit} and is always
below one per cent for the N100 model. Similar accuracies have been obtained
for all other models as well. However, despite the goodness of fit, the time of
explosion cannot be accurately reconstructed in most cases as illustrated in
Fig.~\ref{fig:deltat_summary}. The discrepancies between the actual and assumed
functional form of the early light curve evolution are absorbed in the offset
between real and reconstructed explosion time.
Overall, the determination of the explosion time is more accurate and reliable
the earlier the observations are available (see also discussion in
Section~\ref{sec:limitations}). Furthermore and as expected, the explosion time
can be quite well determined for the deflagration models, which follow a power
law.  However, since the power-law index for these models is overall lowest and
smaller than $p=2$ (see Section~\ref{sec:power_law_rise}), the fireball fit
yields the largest deviations for these models. In summary, this exploration
clearly demonstrates that obtaining an accurate power-law fit to observed
early-time photometric data roughly available in one day intervals does not
necessarily imply that the explosion time can be well reconstructed.  This
severely affects constraining progenitor radii from early-time light curves
\citep[e.g.][]{Bloom2012}, which requires an accurate knowledge of the
explosion date. 

A similar problem has also been pointed out by \citet{Hachinger2013} and
\citet{Mazzali2014} who found a discrepancy of $\sim \SI{1}{d}$ between
explosion dates inferred from early-time photometry and spectral modelling for
the Type Ia SNe 2010jn and 2011fe, respectively. This has been interpreted as a
sign of a `dark phase' between the rise of the radioactively powered light
curve and the actual explosion time, which is theoretically expected since it
takes some time until radiation diffuses from the radioactively heated layers
to the surface of the ejecta \citep{Piro2014}. 

\begin{figure}
  \centering
  \includegraphics{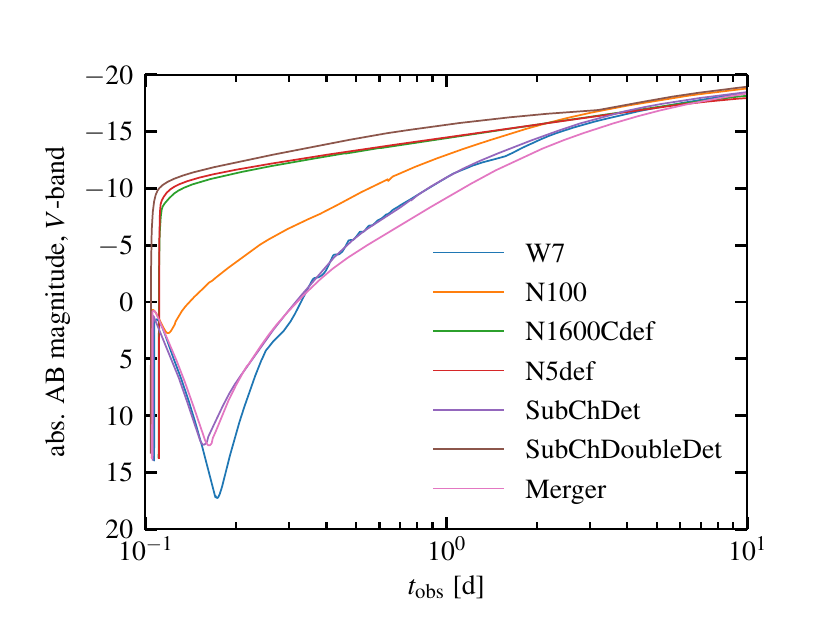}
  \caption{Models with a strongly layered ejecta composition show a delayed
    rise of the radioactively powered light curve. At very early epochs their
    emission is dominated by cooling of the shock-heated ejecta which leads to
    an initial decline in the light curve. Only when radiation from the
  radioactively heated layers starts to diffuse out of the ejecta, the light
curve rises again.}
  \label{fig:dark_phase}
\end{figure}

In our simulations we do find a delay in the rise of the radioactively powered
light curve of $\lesssim \SI{0.2}{d}$ (see Fig.~\ref{fig:dark_phase}).
The actual value for the different models depends strongly on the \nuc{56}{Ni}
distribution in the ejecta (compare also Section~\ref{sec:rise}). Regardless,
the delay associated with this `dark phase' is significantly smaller than the
typical error in the explosion date derived from a power-law fit to the model
light curves. However, we caution that the detailed properties of the `dark
phase', in particular its duration and the absolute magnitude at which the
cooling curve joins the radioactivity-dominated rise, are very sensitive to the
gas temperature in the outer ejecta regions at the start of the calculations.
As pointed out in Section~\ref{sec:mapping_stella}, we use $T = \SI{2e3}{K}$ as
a crude estimate. However, the true temperature value may very well be lower or
higher.

\subsection{Limitations}
\label{sec:limitations}

All results presented in this work are based on one-dimensional calculations.
This simplification is appropriate for some explosion models since they are
either produced in one-dimensional explosion calculations (W7) or do not show
significant asymmetries in the multidimensional explosion simulations (N100,
SubChDet). For others, however, significant asymmetries are predicted
(SubChDoubleDet, merger) and the value of spherically averaged one-dimensional
models may be questioned. For example, the double-detonation model by
\citet{Fink2010} was ignited at the pole and the production of heavy elements
and radioisotopes which lie close to the ejecta surface is strongest in this
region. As a consequence, one expects that the surface radioactivity signatures
as predicted in this work are strongest along lines of sight looking on to this
ignition region and weakest when viewed from opposite directions. A
quantitative estimate of the impact of such deviations from spherical symmetry
is challenging and would require dedicated multidimensional calculations.
However, tools which have been designed for this purpose, such as \artis, are
currently ill-suited to treat the optically thick conditions at very early
times.

In addition to the restriction to one-dimensional geometries, \stella
compromises with respect to the level of detail in the radiative transfer
treatment in order to offer an implicit solution to the full radiation
hydrodynamical problem. For instance, ionization and excitation are assumed to
be governed by LTE. This assumption should be appropriate at the very early
phases during which the ejecta are largely optically thick, but it becomes more
and more inaccurate as the ejecta expand.  Despite the LTE assumption for the
calculation of the plasma state, the source function, $S_{\lambda}$, is non-LTE
$\mbox{\citep{Blinnikov2006}}$
\begin{equation}
  S_{\lambda} = \varepsilon_{\mathrm{th}} B_{\lambda} + (1 - \varepsilon_{\mathrm{th}}) J_{\lambda}.
  \label{eq:source_function}
\end{equation}
Nevertheless, based on the findings by \citet{Baron1996}, \stella adopts a
global thermalization parameter $\varepsilon_{\mathrm{th}} = 1$ for line
interactions \citep{Blinnikov2006}, which leads to a simplified treatment of
fluorescence processes. Again, the consequences of all these simplifications
are difficult to assess with the tools at hand. Eventually, dedicated and more
detailed radiative transfer treatments should be used to predict early
observables. In this context, the inefficiency of standard Monte Carlo
approaches in optically thick regimes prohibits a simple extension of \artis
calculations to early times.

\begin{figure}
  \centering
  \includegraphics{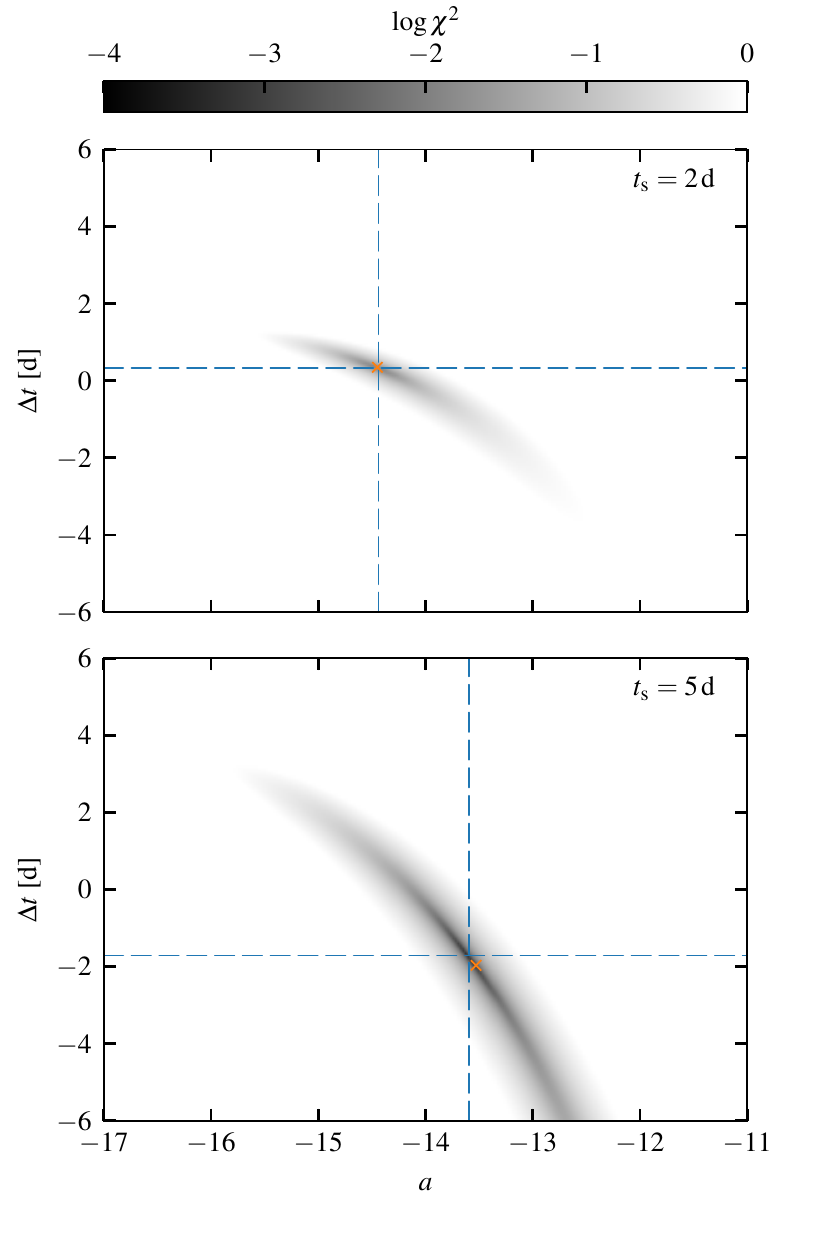}
  \caption{Illustration of the $(a, \Delta t)$ parameter space in the fireball
    fitting process of the N100 model (cf.\ Fig.~\ref{fig:exptime_fit}) for
    $t_{\mathrm{s}} = \SI{2}{d}$ (top panel) and $t_{\mathrm{s}} = \SI{5}{d}$
    (bottom panel). The goodness of fit in terms of the $\chi^2$ value is shown
    in grey-scale for parameter combinations close to the best-fitting values
    reported in Table~\ref{tab:fitfb_params} and included here by dashed blue
    lines. As later data are used for the fit, the region in the parameter
    space corresponding to small $\chi^2$ values increases drastically. As a
    consequence, many more values for $\Delta t$ would result in fits of
    comparable quality and already slight changes in the photometric data can
    result in considerable differences in the obtained fitting parameters. We
    have observed this when repeating the fitting procedure for the N100
    simulation started at $t_{\mathrm{exp}} = \SI{e3}{s}$ (orange crosses).
    While the obtained light curves are virtually identical to the original
    ones, different values for $a$ and in $\Delta t$ are found for
    $t_{\mathrm{s}} = \SI{5}{d}$. These findings emphasize once more that the
    reconstruction of explosion times by means of power-law fits becomes
  increasingly uncertain as ever later data is used for the procedure.}
  \label{fig:N100_fitting_sensitivity_fb}
\end{figure}

For simplicity, all \stella calculations have been started at $t_{\mathrm{exp}}
= \SI{e4}{s}$.  As outlined in Section~\ref{sec:mapping_stella}, the ejecta are
very dense at these epochs and virtually opaque for radiation. Thus we do not
expect a significant contribution to the observables at these times or an
influence on the later evolution. For completeness, we have verified this by
recalculating the N100, W7 and SubChDoubleDet models for $t_{\mathrm{exp}} =
\SI{e3}{s}$. Hereby, we found no significant deviations in the light curves as
shown in Fig.~\ref{fig:earlyrise_summary}. Also the fits performed in
Section~\ref{sec:exp_time} are largely unaffected when starting the simulations
earlier. Only for $t_{\mathrm{s}} = \SI{4}{d}$ and $\SI{5}{d}$, differences can
be observed, prominently in $\Delta t$. However, as detailed in
Section~\ref{sec:exp_time} and shown in Fig.~\ref{fig:deltat_summary}, the
largest deviations between the actual and inferred explosion time, as measured
by $\Delta t$, were already observed in these situations. Also, the fitting
process becomes increasingly uncertain for larger $t_{\mathrm{s}}$. This is
illustrated at the example of the fireball fit to the N100 synthetic light
curve in Fig.~\ref{fig:N100_fitting_sensitivity_fb}.
It shows the behaviour of $\chi^2$ from equation~(\ref{eq:chi2}) in the $(a,
\Delta t)$ parameter space around the best-fitting values. This illustration
highlights that the region of parameter space that corresponds to good-quality
fits (i.e.\ low $\chi^2$) grows as photometric data points further from the
time of explosion are fitted. A very similar behaviour is observed when the
generic power law is used. On the one hand, this explains why slightly
different fitting results are obtained for light curves, which are almost
identical. On the other hand, this finding strongly supports the statement
already made in Section~\ref{sec:exp_time} about the increasing uncertainty of
the explosion time reconstruction for light curve fits which use ever later
data.

For the double detonation, which has radioactive material located close to the
surface, the light curve rise starts slightly earlier when the simulations are
initiated at an earlier time.  In particular, the steep rise at $\sim
\SI{0.12}{d}$ (see insets in in Fig.~\ref{fig:earlyrise_summary}) is superseded
by a more gentle slope. However, the signal is so faint at these early epochs
that this is of little practical consequence. Also the corresponding data point
in Fig.~\ref{eq:column_density} would only shift slightly to earlier times. The
same argument applies to the deflagration models where radioactive material is
spread throughout the ejecta.

We also explored the consequences of our assumption about the initial
isothermal temperature. As discussed in Section~\ref{sec:exp_time}, the
temperature assumption should mainly affect the initial cooling phase. Thus,
different temperatures will quantitatively change the results shown in
Fig.~\ref{fig:dark_phase}. However, the later light-curve evolution that is
dominated by radioactive decay, should be unaffected. We explicitly verified
this at the example of the W7 model, by performing additional calculations with
$T_{\mathrm{iso}} = \SI{e4}{K}$, $\SI{5e4}{K}$ and $\SI{e5}{K}$. In all
calculations, the light curves after the initial cooling phase remain
identical. In particular, the light curve break discussed in
Section~\ref{sec:stratification} and attributed to stratification persists.
Only in the calculation with the highest assumed initial temperature the break
is washed out slightly. This simply occurs because the initial cooling period
now extends into the epoch at which the break happens. Ultimately, we emphasize
that we do not expect high initial temperatures (i.e.\ at $t_{\mathrm{exp}}$)
to prevail in the ejecta due to the strong effect of adiabatic cooling.

\section{Summary and Conclusions}
\label{sec:summary}

With our radiation hydrodynamical calculations presented in this work, we
provide predictions for early light curves for a number of \snia{} explosion
models based on detailed radiative transfer simulations. Compared to previous
works by, for example, \citet{Piro2016}, we perform detailed radiative transfer
simulations and consider explosion models, which are mostly based on
self-consistent, state-of-the art explosion calculations. \citet{Dessart2014b}
use even more sophisticated radiative transfer methods, however, their analysis
does not include the very early epochs ($t\lesssim \SI{1}{d}$) and their sample
is restricted to Chandrasekhar-mass explosions. With our calculations, we aim
at providing predictions for upcoming surveys and at exploring whether
different explosion scenarios may be robustly differentiated based on very
early photometric data.

We have calculated and presented light curves in the $U$, $B$, $V$ and $R$
bands for all our models during the first ten days after explosion. Overall, we
find a very similar light-curve evolution during these early phases for most of
the examined models. Differences are observed in the time when a certain
absolute magnitude is reached, which can be linked to different diffusion times
due to varying amounts of material on top of the \nuc{56}{Ni}-rich zones.

Notwithstanding the similar light-curve evolution, a number of ejecta
properties were identified that imprint characteristic signatures on to the
early observables. Most prominently, radioactive material close to the surface,
as present in the double-detonation model, leads to a fast and very early rise
of the light curve and extended shoulders or even a first maximum. However, a
very similar behaviour is predicted for the interaction between ejecta and a
companion \citep{Kasen2010} or between ejecta and CSM \citep{Piro2016}. Thus,
it may prove challenging to identify the true underlying cause of such
signatures from an observational point of view. Additional information, for
example in terms of the detailed spectral behaviour or the emission in more
energetic regimes, may be required to differentiate between these
possibilities. For example, the detection of narrow or intermediate-width
emission lines in early spectra would be strong indications for ejecta--CSM
interaction. Likewise, the observation of $\gamma$-lines early on may point
towards radioactive material close to the surface. However, such observations
are only realistic for very nearby objects.  In the past, this was possible in
the case of SN~2014J, for which narrow $\gamma$ lines were detected and
attributed to \nuc{56}{Ni} \citep{Diehl2014} located in the outer ejecta
regions. 

Finally, we have identified a modest break in early light-curve evolution in
the W7 model and demonstrated that this feature may be a generic consequence of
strong stratification in the supernova ejecta. The opposite situation of
complete mixing in the ejecta, as seen in the pure deflagration models we
investigated, also impacts the early light curves and leads to a strict
power-law rise. Since \sniasx{} are currently strongly associated with weak
deflagrations \citep{Kromer2013}, we predict that the early light curves of
such objects should feature a power-law rise.

Such a power-law rise of the light curve is often assumed when attempting to
determine explosion times from early observational data. However, apart from
the completely mixed deflagration models, none of the investigated models
yielded light curves that evolved in such a manner. This has important
consequences for reconstructing explosion times. We have illustrated this by
fitting power laws to observations generated from our synthetic light curves.
Despite the non-power-law evolution of the underlying models, we were able to
fit the data with generic power laws and the fireball model to high accuracy,
with deviations less than one per cent. This agreement was achieved, however,
by a very inaccurate dating of the explosion time. In our tests, deviations of
one to two days were very commonly encountered.

Upcoming survey programmes will establish a vast pool of very early data for
\snias{}. We demonstrated with our calculations, that based on early
photometric data alone, it is challenging to unambiguously identify certain
explosion scenarios or specific properties or mechanisms. In this context, our
calculations should be viewed as first explorations, which should be expanded
in the future by detailed predictions about the early spectral evolution and
appearance in the $\gamma$- or X-ray regimes.

\section*{Acknowledgements}

The authors thank R.~Pakmor and S.~Sim for many invaluable discussions,
B.~Shappee for fruitful correspondence and the anonymous reviewer for valuable
comments, which helped to improve the paper. This work has been supported by
the Transregional Collaborative Research Center TRR33 `The Dark Universe' of
the Deutsche Forschungsgemeinschaft and by the Cluster of Excellence `Origin
and Structure of the Universe' at Munich Technical University. The work of ES
(statistical approach for opacity calculation) and PB (inclusion of new
radioactive decay chains as additional sources of $\gamma$-energy) has been
supported by a grant of the Russian Science Foundation 16-12-10519. The Swiss
National Science Foundation has supported the work of SB through Grant no.\
IZ73Z0 152485 SCOPES. MK acknowledges support from the Klaus Tschira
Foundation.  PB is grateful to W.~Hillebrandt and U.~Noebauer for hospitality
during his stay at the MPA. Data analysis and visualization was done using
\textsc{numpy}, \textsc{scipy} \citep{Walt2011} and \textsc{matplotlib}
\citep{Hunter2007}. 




\begin{thebibliography}{}
\makeatletter
\relax
\def\mn@urlcharsother{\let\do\@makeother \do\$\do\&\do\#\do\^\do\_\do\%\do\~}
\def\mn@doi{\begingroup\mn@urlcharsother \@ifnextchar [ {\mn@doi@}
  {\mn@doi@[]}}
\def\mn@doi@[#1]#2{\def\@tempa{#1}\ifx\@tempa\@empty \href
  {http://dx.doi.org/#2} {doi:#2}\else \href {http://dx.doi.org/#2} {#1}\fi
  \endgroup}
\def\mn@eprint#1#2{\mn@eprint@#1:#2::\@nil}
\def\mn@eprint@arXiv#1{\href {http://arxiv.org/abs/#1} {{\tt arXiv:#1}}}
\def\mn@eprint@dblp#1{\href {http://dblp.uni-trier.de/rec/bibtex/#1.xml}
  {dblp:#1}}
\def\mn@eprint@#1:#2:#3:#4\@nil{\def\@tempa {#1}\def\@tempb {#2}\def\@tempc
  {#3}\ifx \@tempc \@empty \let \@tempc \@tempb \let \@tempb \@tempa \fi \ifx
  \@tempb \@empty \def\@tempb {arXiv}\fi \@ifundefined
  {mn@eprint@\@tempb}{\@tempb:\@tempc}{\expandafter \expandafter \csname
  mn@eprint@\@tempb\endcsname \expandafter{\@tempc}}}

\bibitem[\protect\citeauthoryear{{Baklanov}, {Sorokina}  \&
  {Blinnikov}}{{Baklanov} et~al.}{2015}]{Baklanov2015}
{Baklanov} P.~V.,  {Sorokina} E.~I.,   {Blinnikov} S.~I.,  2015, \mn@doi
  [Astronomy Letters] {10.1134/S1063773715040027}, \href
  {http://ads.ari.uni-heidelberg.de/abs/2015AstL...41...95B} {41, 95}

\bibitem[\protect\citeauthoryear{{Baron}, {Hauschildt}, {Nugent}  \&
  {Branch}}{{Baron} et~al.}{1996}]{Baron1996}
{Baron} E.,  {Hauschildt} P.~H.,  {Nugent} P.,   {Branch} D.,  1996, \mn@doi
  [\mnras] {10.1093/mnras/283.1.297}, \href
  {http://ads.nao.ac.jp/abs/1996MNRAS.283..297B} {283, 297}

\bibitem[\protect\citeauthoryear{{Baron}, {Bongard}, {Branch}  \&
  {Hauschildt}}{{Baron} et~al.}{2006}]{Baron2006}
{Baron} E.,  {Bongard} S.,  {Branch} D.,   {Hauschildt} P.~H.,  2006, \mn@doi
  [\apj] {10.1086/504101}, \href {http://ads.nao.ac.jp/abs/2006ApJ...645..480B}
  {645, 480}

\bibitem[\protect\citeauthoryear{{Bessell} \& {Murphy}}{{Bessell} \&
  {Murphy}}{2012}]{Bessell2012}
{Bessell} M.,  {Murphy} S.,  2012, \mn@doi [\pasp] {10.1086/664083}, \href
  {http://ads.ari.uni-heidelberg.de/abs/2012PASP..124..140B} {124, 140}

\bibitem[\protect\citeauthoryear{{Bianco} et~al.,}{{Bianco}
  et~al.}{2011}]{Bianco2011}
{Bianco} F.~B.,  et~al., 2011, \mn@doi [\apj] {10.1088/0004-637X/741/1/20},
  \href {http://ads.nao.ac.jp/abs/2011ApJ...741...20B} {741, 20}

\bibitem[\protect\citeauthoryear{{Blinnikov} \& {Bartunov}}{{Blinnikov} \&
  {Bartunov}}{1993}]{Blinnikov1993}
{Blinnikov} S.~I.,  {Bartunov} O.~S.,  1993, \aap, \href
  {http://adsabs.harvard.edu/abs/1993A%26A...273..106B} {273, 106}

\bibitem[\protect\citeauthoryear{{Blinnikov} \& {Khokhlov}}{{Blinnikov} \&
  {Khokhlov}}{1986}]{Blinnikov1986}
{Blinnikov} S.~I.,  {Khokhlov} A.~M.,  1986, Soviet Astronomy Letters, \href
  {http://adsabs.harvard.edu/abs/1986SvAL...12..131B} {12, 131}

\bibitem[\protect\citeauthoryear{{Blinnikov} \& {Khokhlov}}{{Blinnikov} \&
  {Khokhlov}}{1987}]{Blinnikov1987}
{Blinnikov} S.~I.,  {Khokhlov} A.~M.,  1987, Soviet Astronomy Letters, \href
  {http://adsabs.harvard.edu/abs/1987SvAL...13..364B} {13, 364}

\bibitem[\protect\citeauthoryear{{Blinnikov} \& {Sorokina}}{{Blinnikov} \&
  {Sorokina}}{2000}]{Blinnikov2000a}
{Blinnikov} S.~I.,  {Sorokina} E.~I.,  2000, \aap, \href
  {http://adsabs.harvard.edu/abs/2000A%26A...356L..30B} {356, L30}

\bibitem[\protect\citeauthoryear{{Blinnikov} \& {Tolstov}}{{Blinnikov} \&
  {Tolstov}}{2011}]{Blinnikov2011}
{Blinnikov} S.~I.,  {Tolstov} A.~G.,  2011, \mn@doi [Astronomy Letters]
  {10.1134/S1063773711010051}, \href
  {http://ads.nao.ac.jp/abs/2011AstL...37..194B} {37, 194}

\bibitem[\protect\citeauthoryear{{Blinnikov}, {Eastman}, {Bartunov},
  {Popolitov}  \& {Woosley}}{{Blinnikov} et~al.}{1998}]{Blinnikov1998}
{Blinnikov} S.~I.,  {Eastman} R.,  {Bartunov} O.~S.,  {Popolitov} V.~A.,
  {Woosley} S.~E.,  1998, \mn@doi [\apj] {10.1086/305375}, \href
  {http://ads.nao.ac.jp/abs/1998ApJ...496..454B} {496, 454}

\bibitem[\protect\citeauthoryear{{Blinnikov}, {Lundqvist}, {Bartunov}, {Nomoto}
   \& {Iwamoto}}{{Blinnikov} et~al.}{2000}]{Blinnikov2000}
{Blinnikov} S.,  {Lundqvist} P.,  {Bartunov} O.,  {Nomoto} K.,   {Iwamoto} K.,
  2000, \mn@doi [\apj] {10.1086/308588}, \href
  {http://ads.nao.ac.jp/abs/2000ApJ...532.1132B} {532, 1132}

\bibitem[\protect\citeauthoryear{{Blinnikov}, {R{\"o}pke}, {Sorokina},
  {Gieseler}, {Reinecke}, {Travaglio}, {Hillebrandt}  \&
  {Stritzinger}}{{Blinnikov} et~al.}{2006}]{Blinnikov2006}
{Blinnikov} S.~I.,  {R{\"o}pke} F.~K.,  {Sorokina} E.~I.,  {Gieseler} M.,
  {Reinecke} M.,  {Travaglio} C.,  {Hillebrandt} W.,   {Stritzinger} M.,  2006,
  \mn@doi [\aap] {10.1051/0004-6361:20054594}, \href
  {http://ads.nao.ac.jp/abs/2006A%26A...453..229B} {453, 229}

\bibitem[\protect\citeauthoryear{{Bloom} et~al.,}{{Bloom}
  et~al.}{2012}]{Bloom2012}
{Bloom} J.~S.,  et~al., 2012, \mn@doi [\apjl] {10.1088/2041-8205/744/2/L17},
  \href {http://ads.nao.ac.jp/abs/2012ApJ...744L..17B} {744, L17}

\bibitem[\protect\citeauthoryear{{Branch}, {Doggett}, {Nomoto}  \&
  {Thielemann}}{{Branch} et~al.}{1985}]{Branch1985}
{Branch} D.,  {Doggett} J.~B.,  {Nomoto} K.,   {Thielemann} F.-K.,  1985,
  \mn@doi [\apj] {10.1086/163329}, \href
  {http://ads.nao.ac.jp/abs/1985ApJ...294..619B} {294, 619}

\bibitem[\protect\citeauthoryear{{Bulla}, {Sim}, {Pakmor}, {Kromer},
  {Taubenberger}, {R{\"o}pke}, {Hillebrandt}  \& {Seitenzahl}}{{Bulla}
  et~al.}{2016}]{Bulla2016}
{Bulla} M.,  {Sim} S.~A.,  {Pakmor} R.,  {Kromer} M.,  {Taubenberger} S.,
  {R{\"o}pke} F.~K.,  {Hillebrandt} W.,   {Seitenzahl} I.~R.,  2016, \mn@doi
  [\mnras] {10.1093/mnras/stv2402}, \href
  {http://ads.nao.ac.jp/abs/2016MNRAS.455.1060B} {455, 1060}

\bibitem[\protect\citeauthoryear{{Cao} et~al.,}{{Cao} et~al.}{2015}]{Cao2015}
{Cao} Y.,  et~al., 2015, \mn@doi [\nat] {10.1038/nature14440}, \href
  {http://ads.nao.ac.jp/abs/2015Natur.521..328C} {521, 328}

\bibitem[\protect\citeauthoryear{{Conley} et~al.,}{{Conley}
  et~al.}{2006}]{Conley2006}
{Conley} A.,  et~al., 2006, \mn@doi [\aj] {10.1086/507788}, \href
  {http://ads.nao.ac.jp/abs/2006AJ....132.1707C} {132, 1707}

\bibitem[\protect\citeauthoryear{{Dessart}, {Blondin}, {Hillier}  \&
  {Khokhlov}}{{Dessart} et~al.}{2014a}]{Dessart2014b}
{Dessart} L.,  {Blondin} S.,  {Hillier} D.~J.,   {Khokhlov} A.,  2014a, \mn@doi
  [\mnras] {10.1093/mnras/stu598}, \href
  {http://ads.nao.ac.jp/abs/2014MNRAS.441..532D} {441, 532}

\bibitem[\protect\citeauthoryear{{Dessart}, {Hillier}, {Blondin}  \&
  {Khokhlov}}{{Dessart} et~al.}{2014b}]{Dessart2014a}
{Dessart} L.,  {Hillier} D.~J.,  {Blondin} S.,   {Khokhlov} A.,  2014b, \mn@doi
  [\mnras] {10.1093/mnras/stu789}, \href
  {http://ads.ari.uni-heidelberg.de/abs/2014MNRAS.441.3249D} {441, 3249}

\bibitem[\protect\citeauthoryear{{Diehl} et~al.,}{{Diehl}
  et~al.}{2014}]{Diehl2014}
{Diehl} R.,  et~al., 2014, \mn@doi [Science] {10.1126/science.1254738}, \href
  {http://ads.nao.ac.jp/abs/2014Sci...345.1162D} {345, 1162}

\bibitem[\protect\citeauthoryear{{Fink}, {R{\"o}pke}, {Hillebrandt},
  {Seitenzahl}, {Sim}  \& {Kromer}}{{Fink} et~al.}{2010}]{Fink2010}
{Fink} M.,  {R{\"o}pke} F.~K.,  {Hillebrandt} W.,  {Seitenzahl} I.~R.,  {Sim}
  S.~A.,   {Kromer} M.,  2010, \mn@doi [\aap] {10.1051/0004-6361/200913892},
  \href {http://ads.nao.ac.jp/abs/2010A%26A...514A..53F} {514, A53}

\bibitem[\protect\citeauthoryear{{Fink} et~al.,}{{Fink}
  et~al.}{2014}]{Fink2014}
{Fink} M.,  et~al., 2014, \mn@doi [\mnras] {10.1093/mnras/stt2315}, \href
  {http://ads.ari.uni-heidelberg.de/abs/2014MNRAS.438.1762F} {438, 1762}

\bibitem[\protect\citeauthoryear{{Firth} et~al.,}{{Firth}
  et~al.}{2015}]{Firth2015}
{Firth} R.~E.,  et~al., 2015, \mn@doi [\mnras] {10.1093/mnras/stu2314}, \href
  {http://ads.nao.ac.jp/abs/2015MNRAS.446.3895F} {446, 3895}

\bibitem[\protect\citeauthoryear{Foley et~al.,}{Foley et~al.}{2013}]{Foley2013}
Foley R.~J.,  et~al., 2013, \mn@doi [\apj] {10.1088/0004-637X/767/1/57}, 767,
  57

\bibitem[\protect\citeauthoryear{{Friend} \& {Abbott}}{{Friend} \&
  {Abbott}}{1986}]{Friend1986}
{Friend} D.~B.,  {Abbott} D.~C.,  1986, \mn@doi [\apj] {10.1086/164809}, \href
  {http://ads.ari.uni-heidelberg.de/abs/1986ApJ...311..701F} {311, 701}

\bibitem[\protect\citeauthoryear{{Gall}, {Taubenberger}, {Kromer}, {Sim},
  {Benetti}, {Blanc}, {Elias-Rosa}  \& {Hillebrandt}}{{Gall}
  et~al.}{2012}]{Gall2012}
{Gall} E.~E.~E.,  {Taubenberger} S.,  {Kromer} M.,  {Sim} S.~A.,  {Benetti} S.,
   {Blanc} G.,  {Elias-Rosa} N.,   {Hillebrandt} W.,  2012, \mn@doi [\mnras]
  {10.1111/j.1365-2966.2012.21999.x}, \href
  {http://ads.nao.ac.jp/abs/2012MNRAS.427..994G} {427, 994}

\bibitem[\protect\citeauthoryear{{Goobar} et~al.,}{{Goobar}
  et~al.}{2015}]{Goobar2015}
{Goobar} A.,  et~al., 2015, \mn@doi [\apj] {10.1088/0004-637X/799/1/106}, \href
  {http://ads.nao.ac.jp/abs/2015ApJ...799..106G} {799, 106}

\bibitem[\protect\citeauthoryear{{Hachinger} et~al.,}{{Hachinger}
  et~al.}{2013}]{Hachinger2013}
{Hachinger} S.,  et~al., 2013, \mn@doi [\mnras] {10.1093/mnras/sts492}, \href
  {http://ads.nao.ac.jp/abs/2013MNRAS.429.2228H} {429, 2228}

\bibitem[\protect\citeauthoryear{{Hayden} et~al.,}{{Hayden}
  et~al.}{2010a}]{Hayden2010}
{Hayden} B.~T.,  et~al., 2010a, \mn@doi [\apj] {10.1088/0004-637X/712/1/350},
  \href {http://ads.nao.ac.jp/abs/2010ApJ...712..350H} {712, 350}

\bibitem[\protect\citeauthoryear{{Hayden} et~al.,}{{Hayden}
  et~al.}{2010b}]{Hayden2010a}
{Hayden} B.~T.,  et~al., 2010b, \mn@doi [\apj] {10.1088/0004-637X/722/2/1691},
  \href {http://ads.nao.ac.jp/abs/2010ApJ...722.1691H} {722, 1691}

\bibitem[\protect\citeauthoryear{{Hillebrandt} \& {Niemeyer}}{{Hillebrandt} \&
  {Niemeyer}}{2000}]{Hillebrandt2000}
{Hillebrandt} W.,  {Niemeyer} J.~C.,  2000, \mn@doi [\araa]
  {10.1146/annurev.astro.38.1.191}, \href
  {http://ads.nao.ac.jp/abs/2000ARA%26A..38..191H} {38, 191}

\bibitem[\protect\citeauthoryear{{Hillebrandt}, {Kromer}, {R{\"o}pke}  \&
  {Ruiter}}{{Hillebrandt} et~al.}{2013}]{Hillebrandt2013}
{Hillebrandt} W.,  {Kromer} M.,  {R{\"o}pke} F.~K.,   {Ruiter} A.~J.,  2013,
  \mn@doi [Frontiers of Physics] {10.1007/s11467-013-0303-2}, \href
  {http://ads.nao.ac.jp/abs/2013FrPhy...8..116H} {8, 116}

\bibitem[\protect\citeauthoryear{{Hillier} \& {Dessart}}{{Hillier} \&
  {Dessart}}{2012}]{Hillier2012}
{Hillier} D.~J.,  {Dessart} L.,  2012, \mn@doi [\mnras]
  {10.1111/j.1365-2966.2012.21192.x}, \href
  {http://ads.nao.ac.jp/abs/2012MNRAS.424..252H} {424, 252}

\bibitem[\protect\citeauthoryear{{H{\"o}flich}}{{H{\"o}flich}}{1995}]{Hoeflich1995}
{H{\"o}flich} P.,  1995, \mn@doi [\apj] {10.1086/175505}, \href
  {http://ads.nao.ac.jp/abs/1995ApJ...443...89H} {443, 89}

\bibitem[\protect\citeauthoryear{{Hoyle} \& {Fowler}}{{Hoyle} \&
  {Fowler}}{1960}]{Hoyle1960}
{Hoyle} F.,  {Fowler} W.~A.,  1960, \mn@doi [\apj] {10.1086/146963}, \href
  {http://ads.nao.ac.jp/abs/1960ApJ...132..565H} {132, 565}

\bibitem[\protect\citeauthoryear{{Hunter}}{{Hunter}}{2007}]{Hunter2007}
{Hunter} J.~D.,  2007, \mn@doi [Computing in Science and Engineering]
  {10.1109/MCSE.2007.55}, \href
  {http://ads.ari.uni-heidelberg.de/abs/2007CSE.....9...90H} {9, 90}

\bibitem[\protect\citeauthoryear{{Iben}, {Nomoto}, {Tornambe}  \&
  {Tutukov}}{{Iben} et~al.}{1987}]{Iben1987}
{Iben} Jr. I.,  {Nomoto} K.,  {Tornambe} A.,   {Tutukov} A.~V.,  1987, \mn@doi
  [\apj] {10.1086/165318}, \href {http://ads.nao.ac.jp/abs/1987ApJ...317..717I}
  {317, 717}

\bibitem[\protect\citeauthoryear{Im, Choi, Yoon, Kim, Ehgamberdiev, Monard  \&
  Sung}{Im et~al.}{2015}]{Im2015}
Im M.,  Choi C.,  Yoon S.-C.,  Kim J.-W.,  Ehgamberdiev S.~A.,  Monard L.
  A.~G.,   Sung H.-I.,  2015, \mn@doi [\apjs] {10.1088/0067-0049/221/1/22},
  221, 22

\bibitem[\protect\citeauthoryear{{Iwamoto}, {Brachwitz}, {Nomoto}, {Kishimoto},
  {Umeda}, {Hix}  \& {Thielemann}}{{Iwamoto} et~al.}{1999}]{Iwamoto1999}
{Iwamoto} K.,  {Brachwitz} F.,  {Nomoto} K.,  {Kishimoto} N.,  {Umeda} H.,
  {Hix} W.~R.,   {Thielemann} F.,  1999, \mn@doi [\apjs] {10.1086/313278},
  \href {http://ads.nao.ac.jp/abs/1999ApJS..125..439I} {125, 439}

\bibitem[\protect\citeauthoryear{{Jeffery}, {Leibundgut}, {Kirshner},
  {Benetti}, {Branch}  \& {Sonneborn}}{{Jeffery} et~al.}{1992}]{Jeffery1992}
{Jeffery} D.~J.,  {Leibundgut} B.,  {Kirshner} R.~P.,  {Benetti} S.,  {Branch}
  D.,   {Sonneborn} G.,  1992, \mn@doi [\apj] {10.1086/171787}, \href
  {http://ads.nao.ac.jp/abs/1992ApJ...397..304J} {397, 304}

\bibitem[\protect\citeauthoryear{Jones, Oliphant, Peterson  et~al.}{Jones
  et~al.}{01  }]{Jones2001}
Jones E.,  Oliphant T.,  Peterson P.,   et~al., 2001--, {SciPy}: Open source
  scientific tools for {Python}, \url {http://www.scipy.org/}

\bibitem[\protect\citeauthoryear{Jordan, Perets, Fisher  \& van Rossum}{Jordan
  et~al.}{2012}]{Jordan2012}
Jordan IV G.~C.,  Perets H.~B.,  Fisher R.~T.,   van Rossum D.~R.,  2012,
  \mn@doi [\apjl] {10.1088/2041-8205/761/2/L23}, 761, L23

\bibitem[\protect\citeauthoryear{{Kasen}}{{Kasen}}{2010}]{Kasen2010}
{Kasen} D.,  2010, \mn@doi [\apj] {10.1088/0004-637X/708/2/1025}, \href
  {http://ads.nao.ac.jp/abs/2010ApJ...708.1025K} {708, 1025}

\bibitem[\protect\citeauthoryear{{Kasen}, {Thomas}  \& {Nugent}}{{Kasen}
  et~al.}{2006}]{Kasen2006}
{Kasen} D.,  {Thomas} R.~C.,   {Nugent} P.,  2006, \mn@doi [\apj]
  {10.1086/506190}, \href {http://ads.nao.ac.jp/abs/2006ApJ...651..366K} {651,
  366}

\bibitem[\protect\citeauthoryear{{Kasen}, {R{\"o}pke}  \& {Woosley}}{{Kasen}
  et~al.}{2009}]{Kasen2009}
{Kasen} D.,  {R{\"o}pke} F.~K.,   {Woosley} S.~E.,  2009, \mn@doi [\nat]
  {10.1038/nature08256}, \href {http://ads.nao.ac.jp/abs/2009Natur.460..869K}
  {460, 869}

\bibitem[\protect\citeauthoryear{{Khokhlov}}{{Khokhlov}}{1991}]{Khokhlov1991}
{Khokhlov} A.~M.,  1991, \aap, \href
  {http://adsabs.harvard.edu/abs/1991A%26A...245..114K} {245, 114}

\bibitem[\protect\citeauthoryear{{Kozyreva} et~al.,}{{Kozyreva}
  et~al.}{2017}]{Kozyreva2017}
{Kozyreva} A.,  et~al., 2017, \mn@doi [\mnras] {10.1093/mnras/stw2562}, \href
  {http://ads.nao.ac.jp/abs/2017MNRAS.464.2854K} {464, 2854}

\bibitem[\protect\citeauthoryear{{Kromer} \& {Sim}}{{Kromer} \&
  {Sim}}{2009}]{Kromer2009}
{Kromer} M.,  {Sim} S.~A.,  2009, \mn@doi [\mnras]
  {10.1111/j.1365-2966.2009.15256.x}, \href
  {http://ads.nao.ac.jp/abs/2009MNRAS.398.1809K} {398, 1809}

\bibitem[\protect\citeauthoryear{{Kromer}, {Sim}, {Fink}, {R{\"o}pke},
  {Seitenzahl}  \& {Hillebrandt}}{{Kromer} et~al.}{2010}]{Kromer2010}
{Kromer} M.,  {Sim} S.~A.,  {Fink} M.,  {R{\"o}pke} F.~K.,  {Seitenzahl} I.~R.,
    {Hillebrandt} W.,  2010, \mn@doi [\apj] {10.1088/0004-637X/719/2/1067},
  \href {http://ads.nao.ac.jp/abs/2010ApJ...719.1067K} {719, 1067}

\bibitem[\protect\citeauthoryear{{Kromer} et~al.,}{{Kromer}
  et~al.}{2013}]{Kromer2013}
{Kromer} M.,  et~al., 2013, \mn@doi [\mnras] {10.1093/mnras/sts498}, \href
  {http://ads.nao.ac.jp/abs/2013MNRAS.429.2287K} {429, 2287}

\bibitem[\protect\citeauthoryear{Kromer et~al.,}{Kromer
  et~al.}{2015}]{Kromer2015}
Kromer M.,  et~al., 2015, \mn@doi [\mnras] {10.1093/mnras/stv886}, 450, 3045

\bibitem[\protect\citeauthoryear{{Kromer} et~al.,}{{Kromer}
  et~al.}{2016}]{Kromer2016}
{Kromer} M.,  et~al., 2016, \mn@doi [\mnras] {10.1093/mnras/stw962}, \href
  {http://ads.nao.ac.jp/abs/2016MNRAS.459.4428K} {459, 4428}

\bibitem[\protect\citeauthoryear{Kromer, Ohlmann  \& Roepke}{Kromer
  et~al.}{2017}]{Kromer2017}
Kromer M.,  Ohlmann S.~T.,   Roepke F.~K.,  2017, preprint (\mn@eprint {arXiv}
  {1706.09879})

\bibitem[\protect\citeauthoryear{{Kurucz} \& {Bell}}{{Kurucz} \&
  {Bell}}{1995}]{Kurucz1995}
{Kurucz} R.~L.,  {Bell} B.,  1995, {Atomic Line List}.
Kurucz CD-ROM, Cambridge, MA: Smithsonian Astrophysical Observatory

\bibitem[\protect\citeauthoryear{{Lentz}, {Baron}, {Branch}  \&
  {Hauschildt}}{{Lentz} et~al.}{2001}]{Lentz2001}
{Lentz} E.~J.,  {Baron} E.,  {Branch} D.,   {Hauschildt} P.~H.,  2001, \mn@doi
  [\apj] {10.1086/322239}, \href {http://ads.nao.ac.jp/abs/2001ApJ...557..266L}
  {557, 266}

\bibitem[\protect\citeauthoryear{Liu \& Stancliffe}{Liu \&
  Stancliffe}{2016}]{Liu2016}
Liu Z.-W.,  Stancliffe R.~J.,  2016, \mn@doi [\mnras] {10.1093/mnras/stw774},
  459, 1781

\bibitem[\protect\citeauthoryear{{Livio} \& {Riess}}{{Livio} \&
  {Riess}}{2003}]{Livio2003}
{Livio} M.,  {Riess} A.~G.,  2003, \mn@doi [\apjl] {10.1086/378765}, \href
  {http://ads.nao.ac.jp/abs/2003ApJ...594L..93L} {594, L93}

\bibitem[\protect\citeauthoryear{{Maoz}, {Mannucci}  \& {Nelemans}}{{Maoz}
  et~al.}{2014}]{Maoz2014}
{Maoz} D.,  {Mannucci} F.,   {Nelemans} G.,  2014, \mn@doi [\araa]
  {10.1146/annurev-astro-082812-141031}, \href
  {http://ads.nao.ac.jp/abs/2014ARA%26A..52..107M} {52, 107}

\bibitem[\protect\citeauthoryear{{Marion} et~al.,}{{Marion}
  et~al.}{2016}]{Marion2016}
{Marion} G.~H.,  et~al., 2016, \mn@doi [\apj] {10.3847/0004-637X/820/2/92},
  \href {http://ads.nao.ac.jp/abs/2016ApJ...820...92M} {820, 92}

\bibitem[\protect\citeauthoryear{{Mazzali} et~al.,}{{Mazzali}
  et~al.}{2014}]{Mazzali2014}
{Mazzali} P.~A.,  et~al., 2014, \mn@doi [\mnras] {10.1093/mnras/stu077}, \href
  {http://ads.nao.ac.jp/abs/2014MNRAS.439.1959M} {439, 1959}

\bibitem[\protect\citeauthoryear{{Moll}, {Raskin}, {Kasen}  \&
  {Woosley}}{{Moll} et~al.}{2014}]{Moll2014}
{Moll} R.,  {Raskin} C.,  {Kasen} D.,   {Woosley} S.~E.,  2014, \mn@doi [\apj]
  {10.1088/0004-637X/785/2/105}, \href
  {http://ads.ari.uni-heidelberg.de/abs/2014ApJ...785..105M} {785, 105}

\bibitem[\protect\citeauthoryear{{Noebauer}, {Sim}, {Kromer}, {R{\"o}pke}  \&
  {Hillebrandt}}{{Noebauer} et~al.}{2012}]{Noebauer2012}
{Noebauer} U.~M.,  {Sim} S.~A.,  {Kromer} M.,  {R{\"o}pke} F.~K.,
  {Hillebrandt} W.,  2012, \mn@doi [\mnras] {10.1111/j.1365-2966.2012.21600.x},
  \href {http://ads.ari.uni-heidelberg.de/abs/2012MNRAS.425.1430N} {425, 1430}

\bibitem[\protect\citeauthoryear{{Noebauer}, {Taubenberger}, {Blinnikov},
  {Sorokina}  \& {Hillebrandt}}{{Noebauer} et~al.}{2016}]{Noebauer2016}
{Noebauer} U.~M.,  {Taubenberger} S.,  {Blinnikov} S.,  {Sorokina} E.,
  {Hillebrandt} W.,  2016, \mn@doi [\mnras] {10.1093/mnras/stw2197}, \href
  {http://ads.nao.ac.jp/abs/2016MNRAS.463.2972N} {463, 2972}

\bibitem[\protect\citeauthoryear{{Nomoto}, {Thielemann}  \& {Yokoi}}{{Nomoto}
  et~al.}{1984}]{Nomoto1984}
{Nomoto} K.,  {Thielemann} F.,   {Yokoi} K.,  1984, \mn@doi [\apj]
  {10.1086/162639}, \href {http://ads.nao.ac.jp/abs/1984ApJ...286..644N} {286,
  644}

\bibitem[\protect\citeauthoryear{{Nugent}, {Baron}, {Branch}, {Fisher}  \&
  {Hauschildt}}{{Nugent} et~al.}{1997}]{Nugent1997}
{Nugent} P.,  {Baron} E.,  {Branch} D.,  {Fisher} A.,   {Hauschildt} P.~H.,
  1997, \mn@doi [\apj] {10.1086/304459}, \href
  {http://ads.nao.ac.jp/abs/1997ApJ...485..812N} {485, 812}

\bibitem[\protect\citeauthoryear{{Nugent} et~al.,}{{Nugent}
  et~al.}{2011}]{Nugent2011}
{Nugent} P.~E.,  et~al., 2011, \mn@doi [\nat] {10.1038/nature10644}, \href
  {http://ads.nao.ac.jp/abs/2011Natur.480..344N} {480, 344}

\bibitem[\protect\citeauthoryear{{Pakmor}, {Kromer}, {R{\"o}pke}, {Sim},
  {Ruiter}  \& {Hillebrandt}}{{Pakmor} et~al.}{2010}]{Pakmor2010}
{Pakmor} R.,  {Kromer} M.,  {R{\"o}pke} F.~K.,  {Sim} S.~A.,  {Ruiter} A.~J.,
  {Hillebrandt} W.,  2010, \mn@doi [\nat] {10.1038/nature08642}, \href
  {http://ads.nao.ac.jp/abs/2010Natur.463...61P} {463, 61}

\bibitem[\protect\citeauthoryear{{Pakmor}, {Kromer}, {Taubenberger}, {Sim},
  {R{\"o}pke}  \& {Hillebrandt}}{{Pakmor} et~al.}{2012}]{Pakmor2012}
{Pakmor} R.,  {Kromer} M.,  {Taubenberger} S.,  {Sim} S.~A.,  {R{\"o}pke}
  F.~K.,   {Hillebrandt} W.,  2012, \mn@doi [\apjl]
  {10.1088/2041-8205/747/1/L10}, \href
  {http://ads.nao.ac.jp/abs/2012ApJ...747L..10P} {747, L10}

\bibitem[\protect\citeauthoryear{{Pakmor}, {Kromer}, {Taubenberger}  \&
  {Springel}}{{Pakmor} et~al.}{2013}]{Pakmor2013}
{Pakmor} R.,  {Kromer} M.,  {Taubenberger} S.,   {Springel} V.,  2013, \mn@doi
  [\apjl] {10.1088/2041-8205/770/1/L8}, \href
  {http://ads.ari.uni-heidelberg.de/abs/2013ApJ...770L...8P} {770, L8}

\bibitem[\protect\citeauthoryear{{Pereira} et~al.,}{{Pereira}
  et~al.}{2013}]{Pereira2013}
{Pereira} R.,  et~al., 2013, \mn@doi [\aap] {10.1051/0004-6361/201221008},
  \href {http://ads.nao.ac.jp/abs/2013A%26A...554A..27P} {554, A27}

\bibitem[\protect\citeauthoryear{Phillips et~al.,}{Phillips
  et~al.}{2007}]{Phillips2007}
Phillips M.~M.,  et~al., 2007, \mn@doi [\pasp] {10.1086/518372}, 119, 360

\bibitem[\protect\citeauthoryear{{Pinto} \& {Eastman}}{{Pinto} \&
  {Eastman}}{2000}]{Pinto2000}
{Pinto} P.~A.,  {Eastman} R.~G.,  2000, \mn@doi [\apj] {10.1086/308376}, \href
  {http://ads.nao.ac.jp/abs/2000ApJ...530..744P} {530, 744}

\bibitem[\protect\citeauthoryear{{Piro} \& {Morozova}}{{Piro} \&
  {Morozova}}{2016}]{Piro2016}
{Piro} A.~L.,  {Morozova} V.~S.,  2016, \mn@doi [\apj]
  {10.3847/0004-637X/826/1/96}, \href
  {http://ads.nao.ac.jp/abs/2016ApJ...826...96P} {826, 96}

\bibitem[\protect\citeauthoryear{{Piro} \& {Nakar}}{{Piro} \&
  {Nakar}}{2014}]{Piro2014}
{Piro} A.~L.,  {Nakar} E.,  2014, \mn@doi [\apj] {10.1088/0004-637X/784/1/85},
  \href {http://ads.nao.ac.jp/abs/2014ApJ...784...85P} {784, 85}

\bibitem[\protect\citeauthoryear{{Piro}, {Chang}  \& {Weinberg}}{{Piro}
  et~al.}{2010}]{Piro2010}
{Piro} A.~L.,  {Chang} P.,   {Weinberg} N.~N.,  2010, \mn@doi [\apj]
  {10.1088/0004-637X/708/1/598}, \href
  {http://ads.nao.ac.jp/abs/2010ApJ...708..598P} {708, 598}

\bibitem[\protect\citeauthoryear{{Rabinak}, {Livne}  \& {Waxman}}{{Rabinak}
  et~al.}{2012}]{Rabinak2012}
{Rabinak} I.,  {Livne} E.,   {Waxman} E.,  2012, \mn@doi [\apj]
  {10.1088/0004-637X/757/1/35}, \href
  {http://ads.nao.ac.jp/abs/2012ApJ...757...35R} {757, 35}

\bibitem[\protect\citeauthoryear{{Raskin} \& {Kasen}}{{Raskin} \&
  {Kasen}}{2013}]{Raskin2013}
{Raskin} C.,  {Kasen} D.,  2013, \mn@doi [\apj] {10.1088/0004-637X/772/1/1},
  \href {http://ads.nao.ac.jp/abs/2013ApJ...772....1R} {772, 1}

\bibitem[\protect\citeauthoryear{{Riess} et~al.,}{{Riess}
  et~al.}{1999}]{Riess1999}
{Riess} A.~G.,  et~al., 1999, \mn@doi [\aj] {10.1086/301143}, \href
  {http://ads.nao.ac.jp/abs/1999AJ....118.2675R} {118, 2675}

\bibitem[\protect\citeauthoryear{{R{\"o}pke}}{{R{\"o}pke}}{2005}]{Roepke2005}
{R{\"o}pke} F.~K.,  2005, \mn@doi [\aap] {10.1051/0004-6361:20041700}, \href
  {http://adsabs.harvard.edu/abs/2005A%26A...432..969R} {432, 969}

\bibitem[\protect\citeauthoryear{{R{\"o}pke} et~al.,}{{R{\"o}pke}
  et~al.}{2012}]{Roepke2012}
{R{\"o}pke} F.~K.,  et~al., 2012, \mn@doi [\apjl]
  {10.1088/2041-8205/750/1/L19}, \href
  {http://ads.nao.ac.jp/abs/2012ApJ...750L..19R} {750, L19}

\bibitem[\protect\citeauthoryear{{Rosswog}, {Kasen}, {Guillochon}  \&
  {Ramirez-Ruiz}}{{Rosswog} et~al.}{2009}]{Rosswog2009}
{Rosswog} S.,  {Kasen} D.,  {Guillochon} J.,   {Ramirez-Ruiz} E.,  2009,
  \mn@doi [\apjl] {10.1088/0004-637X/705/2/L128}, \href
  {http://ads.nao.ac.jp/abs/2009ApJ...705L.128R} {705, L128}

\bibitem[\protect\citeauthoryear{{Ruiter}, {Belczynski}  \& {Fryer}}{{Ruiter}
  et~al.}{2009}]{Ruiter2009}
{Ruiter} A.~J.,  {Belczynski} K.,   {Fryer} C.,  2009, \mn@doi [\apj]
  {10.1088/0004-637X/699/2/2026}, \href
  {http://ads.nao.ac.jp/abs/2009ApJ...699.2026R} {699, 2026}

\bibitem[\protect\citeauthoryear{{Salvo}, {Cappellaro}, {Mazzali}, {Benetti},
  {Danziger}, {Patat}  \& {Turatto}}{{Salvo} et~al.}{2001}]{Salvo2001}
{Salvo} M.~E.,  {Cappellaro} E.,  {Mazzali} P.~A.,  {Benetti} S.,  {Danziger}
  I.~J.,  {Patat} F.,   {Turatto} M.,  2001, \mn@doi [\mnras]
  {10.1046/j.1365-8711.2001.03995.x}, \href
  {http://ads.nao.ac.jp/abs/2001MNRAS.321..254S} {321, 254}

\bibitem[\protect\citeauthoryear{{Sasdelli} et~al.,}{{Sasdelli}
  et~al.}{2017}]{Sasdelli2017}
{Sasdelli} M.,  et~al., 2017, \mn@doi [\mnras] {10.1093/mnras/stw3323}, \href
  {http://ads.nao.ac.jp/abs/2017MNRAS.466.3784S} {466, 3784}

\bibitem[\protect\citeauthoryear{{Seitenzahl} et~al.,}{{Seitenzahl}
  et~al.}{2013}]{Seitenzahl2013}
{Seitenzahl} I.~R.,  et~al., 2013, \mn@doi [\mnras] {10.1093/mnras/sts402},
  \href {http://ads.nao.ac.jp/abs/2013MNRAS.429.1156S} {429, 1156}

\bibitem[\protect\citeauthoryear{Shappee, Piro, Stanek, Patel, Margutti,
  Lipunov  \& Pogge}{Shappee et~al.}{2016a}]{Shappee2016a}
Shappee B.~J.,  Piro A.~L.,  Stanek K.~Z.,  Patel S.~G.,  Margutti R.~A.,
  Lipunov V.~M.,   Pogge R.~W.,  2016a, preprint (\mn@eprint {arXiv}
  {1610.07601})

\bibitem[\protect\citeauthoryear{Shappee et~al.,}{Shappee
  et~al.}{2016b}]{Shappee2016}
Shappee B.~J.,  et~al., 2016b, \mn@doi [\apj] {10.3847/0004-637X/826/2/144},
  826, 144

\bibitem[\protect\citeauthoryear{{Shen} \& {Moore}}{{Shen} \&
  {Moore}}{2014}]{Shen2014}
{Shen} K.~J.,  {Moore} K.,  2014, \mn@doi [\apj] {10.1088/0004-637X/797/1/46},
  \href {http://ads.nao.ac.jp/abs/2014ApJ...797...46S} {797, 46}

\bibitem[\protect\citeauthoryear{{Shigeyama}, {Nomoto}, {Yamaoka}  \&
  {Thielemann}}{{Shigeyama} et~al.}{1992}]{Shigeyama1992}
{Shigeyama} T.,  {Nomoto} K.,  {Yamaoka} H.,   {Thielemann} F.-K.,  1992,
  \mn@doi [\apjl] {10.1086/186281}, \href
  {http://ads.nao.ac.jp/abs/1992ApJ...386L..13S} {386, L13}

\bibitem[\protect\citeauthoryear{{Sim}, {Proga}, {Miller}, {Long}  \&
  {Turner}}{{Sim} et~al.}{2010}]{Sim2010}
{Sim} S.~A.,  {Proga} D.,  {Miller} L.,  {Long} K.~S.,   {Turner} T.~J.,  2010,
  \mn@doi [\mnras] {10.1111/j.1365-2966.2010.17215.x}, \href
  {http://ads.nao.ac.jp/abs/2010MNRAS.408.1396S} {408, 1396}

\bibitem[\protect\citeauthoryear{{Soker}, {Garc{\'{\i}}a-Berro}  \&
  {Althaus}}{{Soker} et~al.}{2014}]{Soker2014}
{Soker} N.,  {Garc{\'{\i}}a-Berro} E.,   {Althaus} L.~G.,  2014, \mn@doi
  [\mnras] {10.1093/mnrasl/slt142}, \href
  {http://ads.nao.ac.jp/abs/2014MNRAS.437L..66S} {437, L66}

\bibitem[\protect\citeauthoryear{{Sorokina}, {Blinnikov}, {Nomoto}, {Quimby}
  \& {Tolstov}}{{Sorokina} et~al.}{2016}]{Sorokina2016a}
{Sorokina} E.,  {Blinnikov} S.,  {Nomoto} K.,  {Quimby} R.,   {Tolstov} A.,
  2016, \mn@doi [\apj] {10.3847/0004-637X/829/1/17}, \href
  {http://ads.nao.ac.jp/abs/2016ApJ...829...17S} {829, 17}

\bibitem[\protect\citeauthoryear{Storn \& Price}{Storn \&
  Price}{1997}]{Storn1997}
Storn R.,  Price K.,  1997, \mn@doi [Journal of Global Optimization]
  {10.1023/A:1008202821328}, 11, 341

\bibitem[\protect\citeauthoryear{{Strovink}}{{Strovink}}{2007}]{Strovink2007}
{Strovink} M.,  2007, \mn@doi [\apj] {10.1086/523089}, \href
  {http://ads.nao.ac.jp/abs/2007ApJ...671.1084S} {671, 1084}

\bibitem[\protect\citeauthoryear{{Tanikawa}, {Nakasato}, {Sato}, {Nomoto},
  {Maeda}  \& {Hachisu}}{{Tanikawa} et~al.}{2015}]{Tanikawa2015}
{Tanikawa} A.,  {Nakasato} N.,  {Sato} Y.,  {Nomoto} K.,  {Maeda} K.,
  {Hachisu} I.,  2015, \mn@doi [\apj] {10.1088/0004-637X/807/1/40}, \href
  {http://ads.nao.ac.jp/abs/2015ApJ...807...40T} {807, 40}

\bibitem[\protect\citeauthoryear{{Woosley} \& {Kasen}}{{Woosley} \&
  {Kasen}}{2011}]{Woosley2011}
{Woosley} S.~E.,  {Kasen} D.,  2011, \mn@doi [\apj]
  {10.1088/0004-637X/734/1/38}, \href
  {http://ads.nao.ac.jp/abs/2011ApJ...734...38W} {734, 38}

\bibitem[\protect\citeauthoryear{{Woosley}, {Kasen}, {Blinnikov}  \&
  {Sorokina}}{{Woosley} et~al.}{2007}]{Woosley2007}
{Woosley} S.~E.,  {Kasen} D.,  {Blinnikov} S.,   {Sorokina} E.,  2007, \mn@doi
  [\apj] {10.1086/513732}, \href {http://ads.nao.ac.jp/abs/2007ApJ...662..487W}
  {662, 487}

\bibitem[\protect\citeauthoryear{{Zheng} et~al.,}{{Zheng}
  et~al.}{2013}]{Zheng2013}
{Zheng} W.,  et~al., 2013, \mn@doi [\apjl] {10.1088/2041-8205/778/1/L15}, \href
  {http://ads.nao.ac.jp/abs/2013ApJ...778L..15Z} {778, L15}

\bibitem[\protect\citeauthoryear{{Zheng} et~al.,}{{Zheng}
  et~al.}{2014}]{Zheng2014}
{Zheng} W.,  et~al., 2014, \mn@doi [\apjl] {10.1088/2041-8205/783/1/L24}, \href
  {http://ads.nao.ac.jp/abs/2014ApJ...783L..24Z} {783, L24}

\bibitem[\protect\citeauthoryear{van~der Walt, Colbert  \& Varoquaux}{van~der
  Walt et~al.}{2011}]{Walt2011}
van~der Walt S.,  Colbert S.~C.,   Varoquaux G.,  2011, \mn@doi [Computing in
  Science Engineering] {10.1109/MCSE.2011.37}, 13, 22

\makeatother
\end{thebibliography}





\end{document}